# Spin Torque Efficiency and Analytic Error Rate Estimates of Skyrmion Racetrack Memory

D. Suess[1,1*], C. Vogler[2], F. Bruckner[1], F. Slanovc[1], C. Abert[1]

[1]Doppler Laboratory "Advanced Magnetic Sensing and Materials", University of Vienna, Währinger Straße 17, 1090, Vienna.

[2]Physics of Functional Materials, University of Vienna, Währinger Straße 17, 1090, Vienna.

***Abstract:*** In this paper the thermal stability of skyrmion bubbles and the critical currents to move them over pinning sites is investigated. For the used pinning geometries and the used parameters, the unexpected behavior is reported that the energy barrier to overcome the pinning site is larger than the energy barrier of the annihilation of a skyrmion. The annihilation takes place at boundaries by current driven motion as well as due to the excitation over energy barriers, in the absence of currents, without forming Bloch points. It is reported that the pinning sites, which are required to allow thermally stable bits, significantly increase the critical current densities to move the bits in skyrmion like structures to about $j_{crit}$ = 0.62 TA/m$^2$. These currents are similar to those obtained experimentally to move stable skyrmions at room temperature. By calculating the thermal stability as well as the critical current, we can derive the spin torque efficiency $\eta = \Delta E/I_c$ = 0.19 $k_B T_{300}/\mu A$, which is in a similar range to the simulated spin torque efficiency of MRAM structures. Finally, it is shown that the stochastic depinning process of any racetrack like device requires extremely narrow depinning time distribution smaller than ~6% of the current pulse length to reach bit error rates smaller than $10^{-9}$.

## Introduction

Magnetic skyrmions are topological spin structures that have been predicted theoretically [1,2] and were experimentally found in materials with broken inversion symmetry [3–5] or ultrathin films on substrates exhibiting high spin-orbit coupling[6] . In various studies, skyrmions are suggested as a future data storage device due to the small electrical currents to move these topological structures [7–12]. An important property for any storage device is the long-time thermal stability of the stored information. Whereas the transformation of the skyrmion state to the ferromagnetic state (isotropic collapse) is well studied by calculating energy barriers with string like methods [13–16] and direct Langevin dynamics [17] , this stability is only one prerequisite for the overall stability of stored information. Since in Ref ([7–12]) the presence and the absence of a skyrmion represent bit “1” and bit “0” the stability between these two states is the essential feature for the long-time stability. These two states must be separated by an energy barrier which can be realized by pinning centers. In the presenting work, the pinning centers are realized by constrictions in the wire. If in between two constrictions a skyrmion is present the bit “1” is coded, if not a bit “0” is represented. Within this work, we will compare the energy barriers for depinning at well-defined bit positions with boundary annihilation and isotropic annihilation. Furthermore, the spin torque efficiency of these devices is estimated and the expected bit error rate for various depinning time distributions is estimated.

[1]* Correspondence to dieter.suess@univie.ac.at

**Critical current densities**

In order to study the creation and motion of skyrmions a spin drift-diffusion model is used as described by Zhang et al. [18] which is numerically solved using a hybrid finite element / boundary element method [19,20]. In the used model beside the magnetization, the spin accumulation $\boldsymbol{s}$ and the electrical potential $u$ are calculated as function of input currents at the contact leads. Since the timescale of the relaxation of the spin-accumulation is significantly faster than the timescale of the magnetization precession we treat the spin accumulation in equilibrium within each time step of the Landau-Lifshitz Gilbert equation. In the following the magnetization and the spin accumulation are investigated as function of time for the current driven motion of a skyrmion for the geometry as shown in Fig. 1. The structure consists of three leads. The pinned lead is next to a pinned layer with fixed magnetization in order to generate a skyrmion via spin-transfer torque via the spacer layer. Two leads at the very end (front lead, and back lead) are used to drive the created skyrmions. The front and the back lead have the dimensions of 30 nm x 90 nm x 3nm, respectively. The dimensions of the DMI wire are $l$ = 600 nm, $w$ = 90 nm. We model a $[Pt_{1nm}/Co_{0.6nm}/Ta_{1nm}]_3$ multilayer with 3 repetitions that leads to a total Co thickness of $t_{Co}$ = 1.8 nm unless stated differently. Details of the used effective media model are given in the section “Effective media model”.

The diameter of the notches, the spacer layer (thickness $t$ =1.5 nm), the pinned layer (thickness $t$ =9.0 nm) and the pinned lead (thickness $t$ =6.0) is $d$ = 60nm. The maximum mesh size of the tetrahedral mesh is 4.5 nm, which is below the exchange length of 5nm of the considered material. The magnetic parameters of the wire are motivated by recent studies of Co layers on heavy metal layers. In Ref [[21]] $[Pt(3\ nm)/Co(0.9\ nm)/Ta(4\ nm)]_{15}$ multilayers are studied and the following material parameters are reported: anisotropy constant $K_1$ = 0.37 MJ/m$^3$, with the easy axis perpendicular to the film ($z$-axis), saturation polarization $J_s$ = 0.75 T, exchange constant A = 10 pJ/m and the DMI constant is $D$ = 1.5 x $10^{-3}$ J/m². In Ref[29] $(Ir1|Co0.6|Pt1)_{10}|Pt_3$ structures are studied and the following material parameters are used: $K_1$ = 0.72 MJ/m³, $J_s$ = 1.2 T, A = 10 pJ/m, $D$ = 1.9 x $10^{-3}$ J/m². In Ref[22] $K_1$ = 0.8 MJ/m³, $J_s$ = 0.72 T, A = 15 pJ/m, $D$ = 3.0 x $10^{-3}$ J/m².

Within this work we use parameters for the Co layer that are within the range of Ref [21, 22]: $K_1$ = 0.6 MJ/m³, $J_s$ = 0.72 T, A = 15 pJ/m, $D$ = 3 x $10^{-3}$ J/m². For the damping constant we use $\alpha$ = 0.02, the exchange strength between the conducting electrons and the magnetization is $J$ = 4.1x$10^{-20}$ J. Further the dimensionless polarization parameters $\beta$ = 0.9, $\beta'$ = 0.8, the spin-flip relaxation time $\tau_{sf}$ = 5x$10^{-14}$ s and the diffusion constant $D_0$ = $10^{-3}$ m²/s are used. For the following simulations, it should be noted that only the Co like layers are simulated and the thickness of a layer structure only refers to the Co layer thickness. This approach is well justified as discussed in the method section – "Effective media model". The heavy metal layers between the Co layers are not explicitly considered. If current densities are mentioned in the paper, we refer to the current density within the Co layer. The torque that acts on the Co layer is created by the spin accumulation that arises due to magnetization inhomogeneities within the Co layer. Effects of spin-orbit torque (SOT) that may arise due to the heavy metal layers and act on the Co layer are not considered within this study. The spin diffusion length in Pt and Co are reported to be 14 nm and >40 nm, respectively [23]. Hence, the SOT that arises within a heavy metal layer does not only act on the adjacent Co but it will diffuse and act on several Co layers. As a consequence, it is expected that the torque gets partly compensated due to the multilayer structure. For example, a Pt layer in the center of the stack generates a torque within the Co layers that promotes skyrmion motion in the +x direction above the center and in –x-direction for layers below the center. Detailed studies using the spin diffusion model about the compensation and strength of the SOT compared to spin torque is beyond the scope of this work and will be published elsewhere.

The pinned layer which is used to inject a skyrmion has an anisotropy constant of $K_1$ = 3.0 MJ/m³ and the magnetization is antiparallel to the initial magnetization in the DMI wire. In order to nucleate a skyrmion a current is applied between the pinned lead and the back lead. The current density is increased from zero to 1.0 TA/m² within a time of 0.1ns. After 0.08 ns ( j = 0.8 TA/m² ) already a clear formation of the skyrmion can be seen. Due to the almost constant $M_z$ component of the skyrmion within the center, the created object will be named skyrmion bubble. It should be noted that the accurate calculation of critical currents to create skyrmions within the wire will require atomistic discretization due to Bloch points [24] that are formed during creation [25]. In contrast to the creation process of the skyrmion within the wire, the pinning currents and thermal stability over energy barriers at pinning sites or boundaries of the magnet can be well described in a continuous approach as it will be discussed later in detail.

The nucleated skyrmion has magnetization parallel to the pinned layer which leads to pinning of the skyrmion due to the strayfield of the pinned layer. In the following, a current pulse is applied to move the skyrmion into the center of the DMI wire. At this position, no significant strayfield due to the pinned layer is acting on the skyrmion and the critical currents to move the skyrmion over the pinned site can be studied accurately.

In *Fig. 2* the critical current in order to move the skyrmion over the pinning sites is studied. Two different geometrical realizations of the pinning sites are investigated. The diameter of the pinning sites is $d$ = 120 nm (upper structure) and $d$ = 60 nm (lower structure), respectively. The current density rise time is $r$ = 11 GA/(m²ns). The actual current density is given in *Fig. 2* and it is measured in the DMI wire between two notches. The current is applied at the back lead and the potential is fixed to zero at the front lead. The smallest current density that was applied which moves the skyrmion toward and slightly into the pinning site is $j_{crit}$=0.40 TA/m². The minimum current density in experiments to move skyrmions in an Pt/Co/Ta stacks is reported to be $j_{crit}$=0.20 TA/m² Ref[[21]]. Hence, the presented model that does not have any free parameter predicts a critical current that is larger by about a factor of 2-3 compared to the experimental values of skyrmion motion in multilayers that have finite but in detail unknown pinning sites. For weaker pinning sites the simulated current to move skyrmions can be decreased. In addition, it has to be noted that the simulations are performed at zero temperature and the experiments are performed at room temperature. Hence, the simulated critical current should be overestimated compared to room temperature measurements [26,27].

In contrast in Ref[[22]] the Slonczewski term for spin-orbit torque predicts current densities of 0.001 TA/m² to overcome minor pinning sites and a current to overcome the largest pinning sites of 0.05 TA/m² [[22]]. For even larger pinning sites no passing of the skyrmion was possible. Both values are significantly smaller than the experimental data to move skyrmions in multilayers [21]. Good agreement with experimentally obtained critical currents are obtained in disordered films [28] if a spin Hall angle of 0.34 is assumed [29]. However, in multilayers with repeated layer structure[29] it is not obvious if the spin Hall angle can be used that is obtained for one repetition of the multilayer [29] due to diffusion processes between the layers as discussed earlier.

The left column in *Fig. 2* shows simulations where a positive current is applied at the back lead. Here, the skyrmion moves to the left. Interestingly, the critical current density to overcome the pinning site is smaller for the narrow pinning site ($j_{crit}$=0.62 TA/m² for $d$ = 60 nm) compared to the pinning site with $d$ = 120 nm ($j_{crit}$=0.67 TA/m²). This effect is unexpected since the pinning field of a domain wall that is driven by an external field is proportional to the derivative of the cross section area with respect to the moving direction of the domain, $H_p \propto \partial A / \partial x$. This analytically obtained dependence of domain walls is qualitatively in agreement with results of micromagnetic simulations of pinning fields of Ref [30] . Similar results are also reported for critical depinning currents obtained from experiments and simulations in Ref. [[31]]. The origin of larger pinning currents for a larger diameter of the pinning sites may be found in the larger extension of the pinning sites. As a consequence, for the larger radius, the skyrmion cannot move freely between the pinning sites and does not gain velocity during the time the current is ramped up. Detailed analysis of this effect will be an interesting study for future work.

The used model which couples the magnetization dynamics with the spin accumulation self – consistently solves beside the magnetization as function of time also for the electrical potential as function of time as well as the spin accumulation ***s***. The spin accumulation which depends on the magnetization, in turn, acts via the exchange strength *J* between the conducting electrons and magnetization as a torque term on the magnetization. The three components of the spin accumulation are shown in *Fig. 3* for a current strength just before switching over the pinning center as shown by the position of the skyrmion at the bottom of *Fig. 3*. It is important to note that all three components of the spin accumulation are unequal zero and in the same order of magnitude. This is in contrast to the simplified but often used model of Zhang and Li[32] . A similar model is the model of Thiaville et al. [33] that is equivalent to the model of Zhang and Li if the magnitude of the magnetization is constrained to remain constant. The used self-consistent spin drift-diffusion/micromagnetic model is equivalent to the model of Zhang and Li for the limiting case of a vanishing diffusion constant, $D_0 = 0$. The advantage of the Zhang and Li model compared to the used self-consistent spin drift-diffusion/micromagnetic model is the simplified calculation of the spin-transfer torque, which can be expressed explicitly as function of the magnetization. However, the question must be asked if the approximation leads to different results and the combined solution of both magnetization and spin accumulation is required for accurate results.

Hence, in the following, we study the critical current densities in the limit of $D_0 = 0$ as shown in *Fig. 4*. Here, the skyrmion precesses on an elliptical orbit as well as changes its size periodically as function of time. In *Fig. 4* a current density is applied which is slightly smaller than the critical current density to overcome the pinning site.

### Thermal stability and spin torque efficiency

As mentioned in the introduction a prerequisite for any potential storage application is the thermal stability of the stored information. The average lifetime $\tau$ of magnetic states can be expressed within the framework of the transition state theory (TST) as,

$$\tau = \frac{1}{f_0} \exp\left( \frac{-\Delta E}{k_B T} \right) \qquad (1)$$

where, $\Delta E$ is the energy barrier between two stable states and $f_0$ is the attempt frequency. A wide range of attempt frequencies from several Mhz to 10 THz were reported [17,34–39]. In Ref[[38]] a surprisingly high attempt frequency of $f_0 > 10^{17}[Hz]$ was reported as a results of a simulation based on TST. In order to obtain thermally stable bits in hard disk recording an energy barrier of about $\Delta E \geq 40 k_B T_{300}\ to\ 50 k_B T_{300}$ is a common requirement. In the following, we present simulations of the energy barrier using the string method taking into account all energy terms including the DMI interaction [40,41]. As an input of the string method we use the magnetization states of current driven skyrmion motion over one energy barrier as shown in *Fig. 2*. This path leads to the two stable magnetic states on each side of the pinning site as shown in *Fig. 5* by the images indexed with $i$=0 (initial state 1) with $i$ = 19 (initial state 2). The simulation in the right column of *Fig. 5* corresponds to a geometry with $l$ = 500 nm, $w$ = 75 nm, $t$ = 1.5 nm and $d$ = 50nm. Here, all dimensions of the the previous simulations are scaled by a factor 0.83. The reason is that we aim to investigate and present a structure as small as possible which still supports pinning and depinning of a skyrmion without its annihilation. The images indexed $i$ = 2 to $i$ = 18 show magnetic states along the minimum energy path (MEP). This path is the most probable path which is triggered by thermal fluctuation if no external field or current is applied. The energy along this minimum energy path (MEP) is shown in *Fig. 6* (black line) as function of path index $i$. The reason why the stable state $i$ = 0 has lower energy than the state with $i$ = 19 is due to the strayfield of the pinned layer which is magnetized in the direction of the skyrmion and stabilizes the skyrmion. The smaller energy barrier (barrier 2) of the two possible barriers (barrier 1: moving the skyrmion from left to right, barrier 2: moving the skyrmion from right to left) is $\Delta E$ = 16.4 $k_B T_{300}$, which does not provide sufficient thermal stability for stable bits.

The situation becomes even more severe if the dimensions of the structure are scaled to smaller dimensions: $l$ = 450 nm, $w$ = 67.5 nm, $t$ = 1.35 nm and $d$ = 45nm. Here, all spatial coordinates are scaled by a factor *scale size* = 0.75 compared to the original structure of *Fig. 1*. For this structure, the MEP is shown in *Fig. 5* (left column). It can be seen that the skyrmion is annihilated and nucleated again at the boundary along the MEP, which has recently been reported independently [14,15]. It is reasonable that in principle for finite size systems, topologically charged structures may simply be driven out of the sample leading to a homogeneous state with zero topological charge [[42,43]]. The annihilation of a skyrmion at a boundary was already shown during the current driven domain wall motion in *Fig. 2* (bottom right) indicating that this structure can change its topological charge due to current driven motion without forming Bloch points. Different definitions of topological protection are used in the community. According to Ref[[15]] topologically protected means “there is an energy barrier separating the transition of a system from one topological state to another”. According to Ref [[44]] protection means that “the spin configuration cannot be twisted continuously to result in a magnetic configuration with different *S* (for example, a uniformly magnetized one)", where S is the topological charge . According to the wide definition of topological protection of Ref[15] the boundary annihilation is still a topologically protected process since an energy barrier must be overcome. According to the definition of Ref [44] the boundary annihilation is not topologically protected since the skyrmion can be moved continuously to a magnetic configuration with different S as it is also reported in Ref[14]. A further very interesting observation is that the energy barriers in the two investigated structures are significantly different. The smaller structure showing annihilation of the skyrmion via the boundary displays a smaller energy barrier for the same initial path. In order to investigate this effect in more detail the MEP with annihilation, obtained from the simulation with *scale size* = 0.75, is used as the input path for the simulation with *scale size* = 0.83, which originally did not lead to skyrmion annihilation. Hence, we aim to calculate the energy barrier of annihilation for the larger system and compare it with the energy barrier to overcome the pinning site. As a surprising result, the energy barrier for annihilation ($\Delta E = 13\ k_B T_{300}$) is smaller than the minimum energy path over the pinning site ($\Delta E = 17\ k_B T_{300}$) . Interestingly, the initial path of the skyrmion over the pinning site is a stable local minimum energy path. But there exists at least one local minimum energy path (the MEP that shows annihilation of the skyrmion) with an even lower barrier. Most striking about this effect is that the energy barrier at the pinning site is larger than the energy barrier for annihilation. for the investigated number of multilayer repetitions both energy barriers are significantly too small for applications.

In order to study the energy barrier over the pinning site for different geometries of the constriction the lateral dimensions of the wire are scaled and the barrier is calculated. In *Fig. 7*. the energy barrier is shown for structures where the lateral spatial coordinates are scaled by a factor *scale size* compared to the original structure of Fig. 1 with the Co thickness kept constant at $t_{Co}$ = 1.5 nm. Again, a significant reduction of the energy barrier is obtained for *scale size* < 0.75 since the skyrmion is annihilated for this size, despite the initial path guides the skyrmion through the constriction. For *scale size* = 0.83 the energy barrier is significantly higher since the skyrmion is not annihilated. It can be seen that there exists an optimal lateral dimension of the structure for *scale size* = 0.83 which corresponds to *l* = 500 nm, *w* = 75 nm, and *d* = 50nm yielding the highest energy barrier. Hence, for this given film thickness *t* the energy barrier cannot be further increased to make it stable at room temperature.

The annihilation path is via boundary annihilation which is shown in Fig. 8 for a structure with *l* = 400 nm, *w* = 60 nm, *t* = 1.20 nm and *d* = 40 nm. It is interesting to note that the saddle point configuration is a configuration where the entire skyrmion is still well located within the wire. If the skyrmion partly annihilates the energy decreases. As it can be seen in Fig. 8 (*i* = 21) that the spins between the boundary and the skyrmion rotate by 180° to connect the skyrmion core with the boundary. This is a smooth transition of the spins from the saddle point to states in which the skyrmion exits the structure. Clearly, no Bloch points are formed during this annihilation process and the energy barrier of $\Delta E = 7\ k_B T_{300}$ can be well calculated with micromagnetic simulations. Hence, there is no need to perform atomistic simulations or multiscale simulations for this kind of annihilation process at the boundary.

As we have already shown the energy barrier cannot be further increased by changing the pinning site dimensions. One way to increase the energy barrier of the device is to increase the number of repetitions of the layer stack [45].

In the following the energy barriers for three different paths are investigated as function of the total thickness of the Co layer that are (i) the path over the pinning sites from one bit position to the next one, with the initial path taken from a simulation in which the structure is moved by current (see Fig. 5 in the right column) (ii) the path of annihilation, where the skyrmion is annihilated at a pinning site, with the initial path taken from the simulation of Fig. 5 (left) (iii) the path of annihilation, where the skyrmion is annihilated at the top flat edge. Here the initial path is constructed by shifting the position of the skyrmion in the +y direction. It has to be mentioned that a detailed study for the maximum thickness before inhomogeneous sates in *z*-direction are formed will require simulations where the reduced exchange strength between the Co layers due to the Ta and Pt layer are taken into account [46].

For the wire with l = 500 nm, w = 75 nm, d = 50 nm the energy barriers of these three paths are shown by the plots with circles in *Fig. 9*. For both paths that lead to annihilation, the energy barrier increases approximately linearly as function of the film thickness (layer repetitions). For the path over the pinning site the energy barrier does not exactly increase linearly as function of the thickness, which is attributed to a strayfield effect. Since the skyrmion changes size via moving through the pinning site the effect of the strayfield, which changes as function of thickness, becomes important. The simulations indicate that a film thickness of $t_{Co} > 4.8$ nm (8 stack repetitions) is required in order to obtain energy barriers larger than 50 $k_B T_{300}$ for the structure with $l$ = 500 nm, $w$ = 75 nm, $d$ = 50 nm.

For comparison, also the energy barrier for annihilation at a smaller defect ($l$ = 400 nm, $w$ = 60 nm and $d$ = 40 nm) is shown in *Fig. 9*. (black line, rectangles). It can be seen that the energy barrier for the annihilation processes decreases with the pinning diameter. The largest energy barrier for annihilation is obtained at the flat edge.

Finally, we calculate the spin torque efficiency for the structure with *l* = 600 nm, *w* = 90 nm, t = 1.8 nm which can be derived from the energy barrier and the critical current

$\eta = \Delta E/I_c = 0.19\ k_B T_{300}/\mu A$.

Here, it might be important to note that the spin torque efficiency is approximately independent from the film thickness, since both the energy barrier as well as the critical current depend approximately linearly on the film thickness. Hence, this estimate is also valid for thicker films. The spin torque efficiency of a metallic MRAM junction simulated with the same spin diffusion approach is in the similar order, $\eta = \Delta E/I_c = 0.16\ k_B T_{300}/\mu A$ [[47]]. The spin torque efficiency for experimental CoFeB-MgO based tunnel junctions with diameters below 30 nm are in the range between 1 and 10 $k_B T/\mu A$ as reported in Ref [[48]]. Hence, this study indicates that the spin torque efficiency of the investigated skyrmion structures is in the same order of magnitude or smaller compared to MRAM structures if spin transfer torque is the leading torque.

### Bit error rate of race track like devices

Another important topic for any skyrmion like device is the requirement for reliable depinning of the magnetic structure from the pinning center. If current pulses are applied in the investigated skyrmion racetrack like structure the depinning process is a stochastic process due to finite temperature and will have further distribution in time due to imperfections of different pinning centers. The distribution of the depinning time of field-driven and current driven domain walls is investigated in various studies [[49–52]]. For MRAM structures in the dynamical switching regime, the standard deviation of the switching time is in the order of 10% of the switching time [53]. In Ref [[54]] it can be seen that the skyrmion positions are statistically distributed, and after applying current pulses some skyrmions are moving but others do not move due to larger pinning potentials[54]. By means of computer simulations, the distribution of velocities and skyrmion positions are studied for disordered films showing significant challenges for applications [28].

In applications, the current pulse must have a sufficient duration and strength to move the topological structure over the first pinning site as shown in Fig. 10 (c). However, if the current pulse is too long the probability increases that the structure moves too far over the second pinning site Fig. 10 (d). Hence, the optimization of the correct pulse duration is essential to avoid written in errors. Let us assume a current pulse with a particular strength leading to 50% depinning probability at a pulse length $t_a$. Let us further assume that the depinning time is normal distributed with a standard deviation of $\sigma$. The probability that the structure depins within a time $t_{pulse}$ is given by,

$$P_{depinning}(t_a, t_{pulse}, \sigma) = \int_0^{t_{pulse}} f(t_a, t, \sigma)dt = -\frac{1}{2} erf\left(\frac{t_a - t}{\sqrt{2}\sigma}\right)\Bigg|_0^{t_{pulse}} \tag{2}$$

where the probability of a switching event at time *t* within the time interval $dt$ is given by

$$f(t_a, t, \sigma)dt = \frac{1}{\sigma\sqrt{2\pi}} \exp\left[-\frac{1}{2}\left(\frac{t - t_a}{\sigma}\right)^2\right] dt \tag{3}$$

The probability that depinning occurs can be increased if the current pulse duration is increased, but the probability that the topological object moves too far over the next pinning site increases too. The probability that the structure moves over two pinning sites at time *t′* in the time interval $dt'$ is $p_{t'}dt' = \int_0^{t'} f(t_a, t'', \sigma) f(t_a, t' - t'', \sigma) dt'' dt'$ . The total probability that within the time of the pulse $t_{pulse}$ the skyrmion moves over the second pinning site is given by,

$$P_{depinning,2}(t_a, t_{pulse}, \sigma) = \int_0^{t_{pulse}} \int_0^{t'} f(t_a, t'', \sigma) f(t_a, t' - t'', \sigma) dt'' dt' \tag{4}$$

For $t_a \gg \sigma$ The inner integral can be approximated as

$$\int_0^{t'} f(t_a, t'', \sigma) f(t_a, t' - t'', \sigma) dt'' \approx \int_{-\infty}^{\infty} f(t_a, t'', \sigma) f(t_a, t' - t'', \sigma) dt''$$

which leads to

$$P_{depinning,2}(t_a, t_{pulse}, \sigma) \approx P_{depinning}(2t_a, t_{pulse}, \sqrt{2}\sigma) \tag{5}$$

The probability of a successful write process can be expressed as 1 minus the probability that the structure moves across at least two pinning sites minus the probability that the structure does not move across one pinning site. If the racetrack like device consists of *N* bits the before mentioned process must be successfully repeated for *N* times to move the skymrion from the input to the output. Hence, the success probability is given by,

$$\begin{aligned} P_{success} &= \left[1 - P_{depinning,2}(t_a, t_{pulse}, \sigma) - \left(1 - P_{depinning}(t_a, t_{pulse}, \sigma)\right)\right]^N = \\ &= \left[P_{depinning}(t_a, t_{pulse}, \sigma) - P_{depinning,2}(t_a, t_{pulse}, \sigma)\right]^N \end{aligned} \tag{6}$$

*Fig. 11* shows $P_{success}$ as function of the pulse duration $t_{pulse}$ for $\sigma = 0.1$ ns and $t_a$ = 1.0 ns. It can be seen that for too short $t_{pulse}$ the success rate is small since the topological structure cannot move over the first pinning site as shown in Fig. 10 (b). For too long pulses the success rate is small again due to the high probability of moving over two pinning sites. If the pulse duration is $t_{pulse}$ = 1.4 x $t_a$ for $N$=1 the highest success probability $P_{success}$ is obtained. The bit error rate $BER = 1 - P_{success}$ is plotted as function of $\sigma$ in *Fig. 12* for different *N*. For *N* = 1 it shows that increasing sigma from $\sigma = 0.1$ ns to $\sigma = 0.2$ ns changes the success rate significantly from $P_{success}$ = 3.3x$10^{-5}$ to $P_{success}$ = 3.7x$10^{-2}$. To meet the write and the read bit error rate requirements of STT-MRAM which is BER < $10^{-9}$ Ref[55] a standard deviation of the depinning time $\sigma < 0.06$ $t_a$ is required for $N$=10. In principle the time $t_a$ can be increased by increasing the distance of adjacent bits or decreasing the speed of the topological object. Both possibilities lead to unwanted properties for data storage restricting either data rate or data density. Hence, detailed optimizations and accurate estimates of $\sigma$ for example by Langevin simulations are required to access the feasibility of racetrack like devices. A possibility to increases $P_{success}$ might be to apply current pulses with different strengths. First, a small strength just to move the structure to the next pinning site could be used. If the structure is located next to the pinning site a high current pulse to overcome the pinning site can be applied. This clock cycle might lead to an increase of $t_a$ but does not increase $\sigma$.

Finally, we also aim to investigate the requirements for a BER of $10^{-9}$ if not such well defined pinning sites are used as in Fig. 1, but the film consists of a disordered material with random pinning sites[28]. Let us assume the racetrack device stores *N* bits with a center to center distance of the skyrmion positions of *d*. Let us assume that after each current pulse the skyrmion position can be described by a normal distribution with a standard deviation of the new position $\sigma_d$. After *N* pulses a skyrmion moves through the entire structure. If we assume a Gaussian process the standard deviation of the final position is $\sqrt{N}\sigma_d$ . In order to achieve the required bit error rate of $10^{-9}$, six times the final standard deviation of the position must be smaller than the bit length, $6\sqrt{N}\sigma_d < d$ . Hence $\frac{\sigma_d}{d} < \frac{1}{6\sqrt{N}}$. For a Gaussian process, the ratio $\frac{\sigma_d}{d}$ can be decreased with larger *d* according to $\frac{\sigma_{d'}}{d'} = \frac{\sqrt{m}\sigma_d}{md}$, which sacrifices the data density and read/write speed. In order to estimate $\frac{\sigma_d}{d}$ of a skyrmion device we extract the data from Ref[28] where the skyrmion motion at zero temperature in a disordered film is studied by means of simulations. For a current density of $j$ = 0.1 TA/m$^2$ we obtain from Ref[28] *d* = 413 nm and $\sigma_d = 31$ nm, which allows for a maximum skyrmion number of *N* = 4 to meet a BER of $10^{-9}$ in a device with a magnetic wire length of 1.6 μm.

In both examples (disordered film, well-defined pinning sites) concepts that relax the requirements of the distributions, such as structures reported in Ref[56,57], are required.

## Discussion

In this study, the stability of skyrmions, as well as critical currents to move skyrmions, are investigated. Since for a storage application the skyrmion positions must be well defined and thermally stable, skyrmions must be pinned at certain positions. Usually, the presence and the absence of a skyrmion represent the stored information. In this work pinning sites are introduced by geometrical constrictions to define bit positions. Critical currents as well as energy barriers to overcome these pinning sites are investigated. It is shown that the required pinning sites for storage applications lead to critical current densities to write bits that are in a similar range as micromagnetically simulated critical currents in MRAM devices. A very interesting effect is found that the energy barrier to overcome these pinning sites is larger than the energy barrier for skyrmion annihilation in the investigated structure. For the given material parameters, the isotropic annihilation showed the largest energy barrier as discussed in the method section - Minimum energy path. For the investigated multilayer stack more than 7 repetitions are required to obtain sufficiently high barriers. In order to achieve bit error rates as required by MRAM structures, innovative new concepts will be required. Concerning a practical point of view, it is also worth mentioning that bits in hard disks are significantly smaller ( 45 nm x 10 nm ) [58] than the thermally stable skyrmion structures [[54,59]] and the required distances between them reported in this work. As a consequence, applications and markets need to be found in which the size of the magnetic bit is not the leading objective and multiple bits per cell in RAM like devices lead to an increased performance [60].

The financial support by the Austrian Federal Ministry of Science, Research and Economy and the National Foundation for Research, Technology and Development as well as the Austrian Science Fund (FWF) under Grant Nos. F4112 SFB ViCoM, I2214-N20 and the Vienna Science and Technology Fund (WWTF) under Grant No. MA14-044 is acknowledged.

**Methods**

**Solution of the coupled spin drift-diffusion equation and micromagnetic equation**

The discretization of the continuum model is performed using a hybrid finite element / boundary element method as presented in detail in Ref[19]. The total energy of the magnetic system is composed of the exchange energy, the magnetostatic energy, the anisotropy energy and the Zeeman energy [61]. In order to take into account the interface DMI the following energy contribution is added to the total energy $E_{tot}$ [62],

$$E_{DMI} = \int D\left[\mathbf{m} \cdot \nabla m_n - m_n \nabla \mathbf{m}\right] dV \tag{7}$$

where *D* is the DMI constant and $m_n$ is the projected magnetization on the interface normal, defined by $m_n = \mathbf{m} \cdot \mathbf{e}_z$ with $\mathbf{e}_z$ being the normal vector of the interface. For the solution of the equation of the magnetization dynamics, the effective field $\boldsymbol{H}_{eff}$ is calculated on each node point of the finite element mesh with test functions $\boldsymbol{v}$.

$$-\int_\Omega J_s \mathbf{H}_{eff} \mathbf{v} dV = \delta E_{tot}\left(\mathbf{m}, \mathbf{v}\right) \tag{8}$$

The right-hand side of Eq. (8) is the Gâteaux derivative of the total energy $E_{tot}\left(\mathbf{m}\right)$ in the direction $\mathbf{v}$. Mass lumping is used in order to calculate the effective field $\boldsymbol{H}_{eff}$ on the node points from the right-hand side of Eq. (8). The approach using the Gâteaux derivative of the total energy has the advantage that the boundary conditions for the corresponding field terms are considered in a natural fashion and no explicit treatment is required. In contrast, if the effective field on the node points is calculated by solving

$$-\int_\Omega J_s \mathbf{H}_{eff} \mathbf{v} dV = \int_\Omega 2D\left[\nabla\left(\mathbf{e}_z \cdot \mathbf{m}\right) - \mathbf{e}_z\left(\nabla \cdot \mathbf{m}\right)\right] \mathbf{v} dV \tag{9}$$

where the second term is derived from the functional derivative of the total energy

$$\mathbf{H}_{DMI} = -\frac{\delta E_{DMI}}{\delta \mathbf{J}} = -\frac{2D}{J_s}\left[\nabla\left(\mathbf{e}_z \cdot \mathbf{m}\right) - \mathbf{e}_z\left(\nabla \cdot \mathbf{m}\right)\right] \tag{10}$$

the boundary conditions for the normal derivative of the magnetization at the boundary due to the DMI interaction and the exchange interaction,

$$2A\frac{\partial \mathbf{m}}{\partial n}+D\left(\mathbf{e}_z\times\mathbf{n}\right)\times\mathbf{m}=0 \qquad (11)$$

must be explicitly considered.

**Effective media model**

In the following, we will describe the methodology that is used to model the magnetic multilayers. Physically the multilayers consist of magnetic layers with different magnetic properties. As it is common in micromagnetic simulations atomic features are only considered in a continuum sense (for example the layer structures of $L_{10}$ layer materials are not resolved in micromagnetic simulations).

In the following, we start from the magnetic properties of two magnetic layers as shown in Fig. 13 (left). The layers denoted with the label 2 may represent non-magnetic layers such as Ta or Pt. The layer with label 1 may denote the magnetic Co layer. In the following, we aim to approximate the layered model (left) with an effective media model (right) with appropriate parameters.

Let us start to obtain an effective saturation magnetization $J_{s,\mathrm{eff}}$ for the effective media model. In order to obtain the same magnetic moment $\mu$ as in the layered model one requires:

$$\mu = J_{s,1}V_1 + J_{s,2}V_2 = J_{s,eff}V \qquad (12)$$

Hence, one gets:

$$J_{s,eff} = \frac{1}{V}\left(J_{s,1}V_1 + J_{s,2}V_2\right) \qquad (13)$$

Hence, the Zeeman energy is the same for these two models independent from the direction of the magnetization.

In order to derive an effective anisotropy constant, $K_{1,eff}$ it is worth noting, that this cannot simply be obtained by the same argument as used above. Let us derive $K_{1,eff}$ by calculating the total energy in parallel and antiparallel direction, which must be the same for the two models. Since the Zeeman energy and the exchange energy are the same for both configurations these contributions are not included in the following calculation.

$$\left(K_{1,eff}-\frac{1}{2}\frac{J_{s,eff}^{\;2}}{\mu_0}\right)V = E_\perp - E_\parallel = K_{1,1}V_1 - \frac{1}{2}\frac{J_{s,1}^{\;2}}{\mu_0}V_1 + K_{1,2}V_2 - \frac{1}{2}\frac{J_{s,2}^{\;2}}{\mu_0}V_2 \qquad (14)$$

Here, the second term on the left-hand side considers the shape anisotropy of the effective media model. The corresponding shape anisotropies of the two layers in the layered model are considered by the second and fourth term on the right-hand side of Eq. (14). Since, we assume that the film thickness is significantly smaller than the width of the layers the strayfield of one layer to the adjacent layer can be neglected. As a consequence, the layers do not interact which each other due to strayfield interactions and the demagnetizing field must be considered separately for each layer. $K_{1,eff}$ is then simply obtained by

$$K_{1,eff} = \frac{1}{V}\left( K_{1,1}V_1 - \frac{1}{2}\frac{J_{s,1}{}^2}{\mu_0}V_1 + K_{1,2}V_2 - \frac{1}{2}\frac{J_{s,2}{}^2}{\mu_0}V_2 \right) + \frac{1}{2}\frac{J_{s,eff}{}^2}{\mu_0} \tag{15}$$

For the exchange energy, the simple average approach is used as

$$A_{eff} = \frac{1}{V}\left( A_1V_1 + A_2V_2 \right) \tag{16}$$

Here, it should be noted that the approximation of Eq. (16) is well justified for the exchange energy contribution in the direction within the plane. For variations of the magnetization in z-direction (perpendicular direction), due to the different exchange constants within the layers, this approximation will not be well suited. However, within this work the film thickness is thinner than the domain wall width in z-direction and hence the approximation will be well suited since no inhomogeneities in z-direction are expected.

In Table 1 the used material parameters are summarized for a magnetic layer (1) with the material parameters given by column (c). The non-magnetic layer (2) has a thickness of 2 nm, which can be 1 nm Pt and 1 nm Ta. In the layer (2) all magnetic properties are assigned to have value zero. 6 repetitions of the layer stack are assumed. Column (a) denotes the resulting material parameters of the effective media model (b) The media model where only the magnetic layer is simulated and stacked above each other.

Table 1: Material parameters for different nearly equivalent models.

| | Effective model (a) | 6 con. Co layers (b) | 6 sep. Co layers (c) |
|---|---|---|---|
| $K_1$ (MJ/m³) | 0.1 | 0.6 | 0.6 |
| $J_s$ (T) | 0.13 | 0.73 | 0.73 |
| $A$ (A/m) | $3.45 \times 10^{-12}$ | $1.5 \times 10^{-11}$ | $1.5 \times 10^{-11}$ |
| $D$ (mJ/m²) | 0.7 | 3.0 | 3.0 |
| Layer thickness (nm) | 15.6 | 3.6 | 0.6 nm each layer separated by 2 nm air |

The validity of the effective media model is presented in *Fig. 14* and *Fig. 15*. In *Fig. 14* it is shown that the hard axis loop of an extended but finite film agrees very well for the three investigated models. The model (a) is the model of the effective media. Model (b) is a model where only the Co layers are simulated and stacked directly in contact above each other. Since no domain walls are expected within the film thickness this is a good approximation of the structure. Model (c) is a model where 6 Co layers are simulated with air in between, that represents the non-magnetic Ta and Pt layers.

The energy barriers of these three models are compared in *Fig. 15*. Whereas, the barriers of model (a) and model (b) perfectly agree, the model with the 6 Co layers that are not exchange coupled shows a significant difference. For model (c) the skyrmion gets annihilated at the boundary. The origin may be in the unrealistic weak coupling between the layers in the model that is only triggered by strayfield interactions and may lead to an individual motion of the skyrmions within the layers. In reality, we expect a significant exchange coupling between the Co layers. In Ref[[46]] the coupling strength in Co(4Å)/Pt(t PtÅ)/$[Co(4Å)/Pt(6Å)]_2$ as a function of the Pt coupling layer is investigated. For a spacer layer of 2.5 nm a coupling strength of 0.5 $mJ/m^2$ is reported, which results in an exchange constant of $A = 0.6$ $pJ/m^3$.

**Minimum energy path**

For the calculation of the minimum energy path, the string method is used [40,41]. An advantage of the string method (details of the implementation for magnetic systems can be found in Ref[[16]] ) over the nudged elastic band[63,64] method is that it does not require any modifications on the existing micromagnetic simulator. The entire method was implemented within the python interface of the micromagnetic software magnum.fe [65]. In order to relax the images of the string method, the Landau-Lifshitz-Gilbert equation is integrated without the gyromagnetic precession term for 20 ps. After this time the images are equally distributed along the string. As stopping criterion, the change of the energy barrier is used which must be smaller than $10^{-23}$ J. In Ref[[66]] two completely independent atomistic implementations of the barrier for isotropic skyrmion annihilation are implemented. The barrier obtained by the GNEB method ($\Delta E = 4.416 \times 10^{-20}$ J) agrees well with the simple string method ($\Delta E = 4.421 \times 10^{-20}$ J).

The energy barrier as function of the mesh size is shown in Fig. 16 for boundary annihilation and isotropic annihilation via a Bloch point for a system of size 90 nm x 90 nm x 0.6 nm, which represents one 0.6 nm thick Co layer (1 repetition of the layer stack). It can be seen that the boundary annihilation barrier does not depend on the mesh size since no Bloch point is formed. Hence it can be reliably calculated with micromagnetics. As expected, for the isotropic annihilation a mesh size dependence is obtained due to the formation of a Bloch point. However, it can be seen that for mesh sizes reaching the atomistic lattice constant the energy barrier for isotropic annihilation is significantly larger than for boundary annihilation. A detailed comparison between atomistic simulations and finite difference simulations as function of the mesh size shows that for a mesh size equal to the atomistic lattice constant the micromagnetic simulation overestimates the energy barrier by about a factor of two [66].

**Author Contributions**

D.S. performed the simulations and wrote the main manuscript. C.A wrote the original code of the spin drift-diffusion equation coupled to micromagnetic equations. F. B. and C. V. improved the used simulation code and set up initial simulations of skyrmions. F. S. supported the error rate calculations. All authors reviewed and improved the manuscript.

**Additional Information**

Competing financial interests: The authors declare no competing financial interests.

Fig. 1: Geometry of the used structure. The front and the back lead have dimensions of 30 nm x 90 nm x 3nm. The dimensions of the DMI wire are *l* = 600 nm, *w* = 90 nm, *t* = 1.8 nm. The diameter of the notches as well of the spacer layer (*t* =1.5 nm), the pinned layer (*t* =9.0 nm) and the pinned lead (*t* =6.0) is *d* = 60nm.

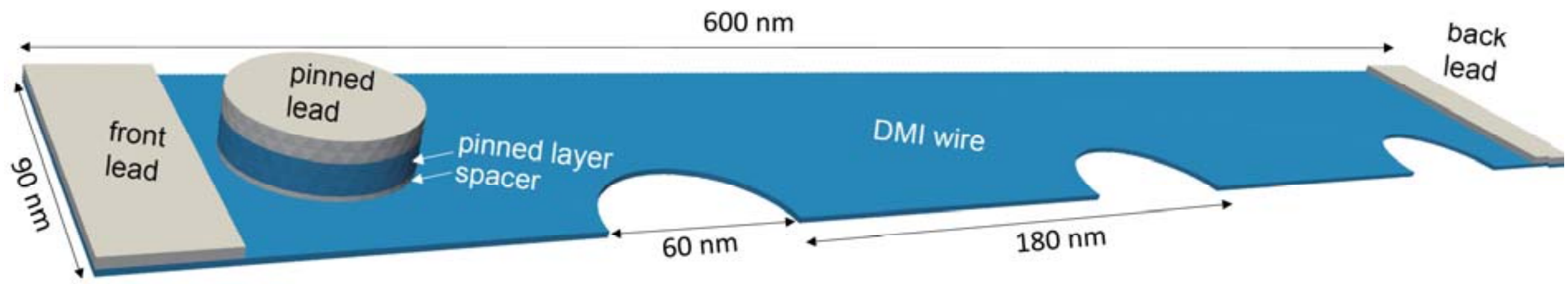

Fig. 2: Motion of the skyrmion as function of the applied current density. The input current is applied at the front lead. The voltage is zero at the back lead. The dimensions of the wire are: *l* = 600 nm, *w* = 90 nm, *t* = 1.8 nm. The diameter of the notches is d = 60nm and d=120 nm, respectively. The given current density is the average density within the wire holding the skyrmion at the center between notches. (left column) a positive current density is applied at the right lead (right column) a negative current is applied at the right lead.

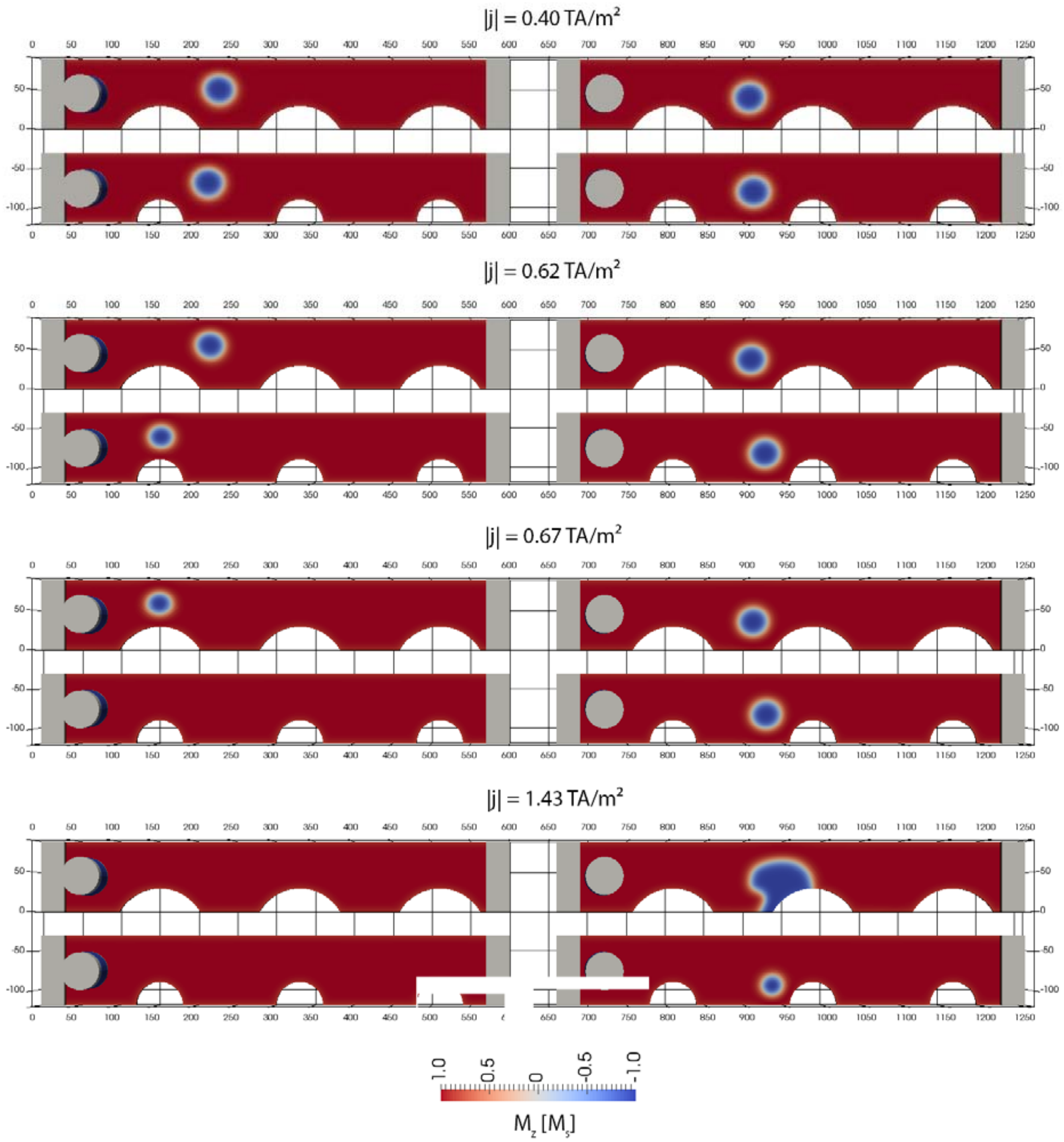

Fig. 3: *x*,*y* and *z* component of the spin-accumulation and magnetization before overcoming the pinning center. The lateral dimensions are as in Fig. 1.

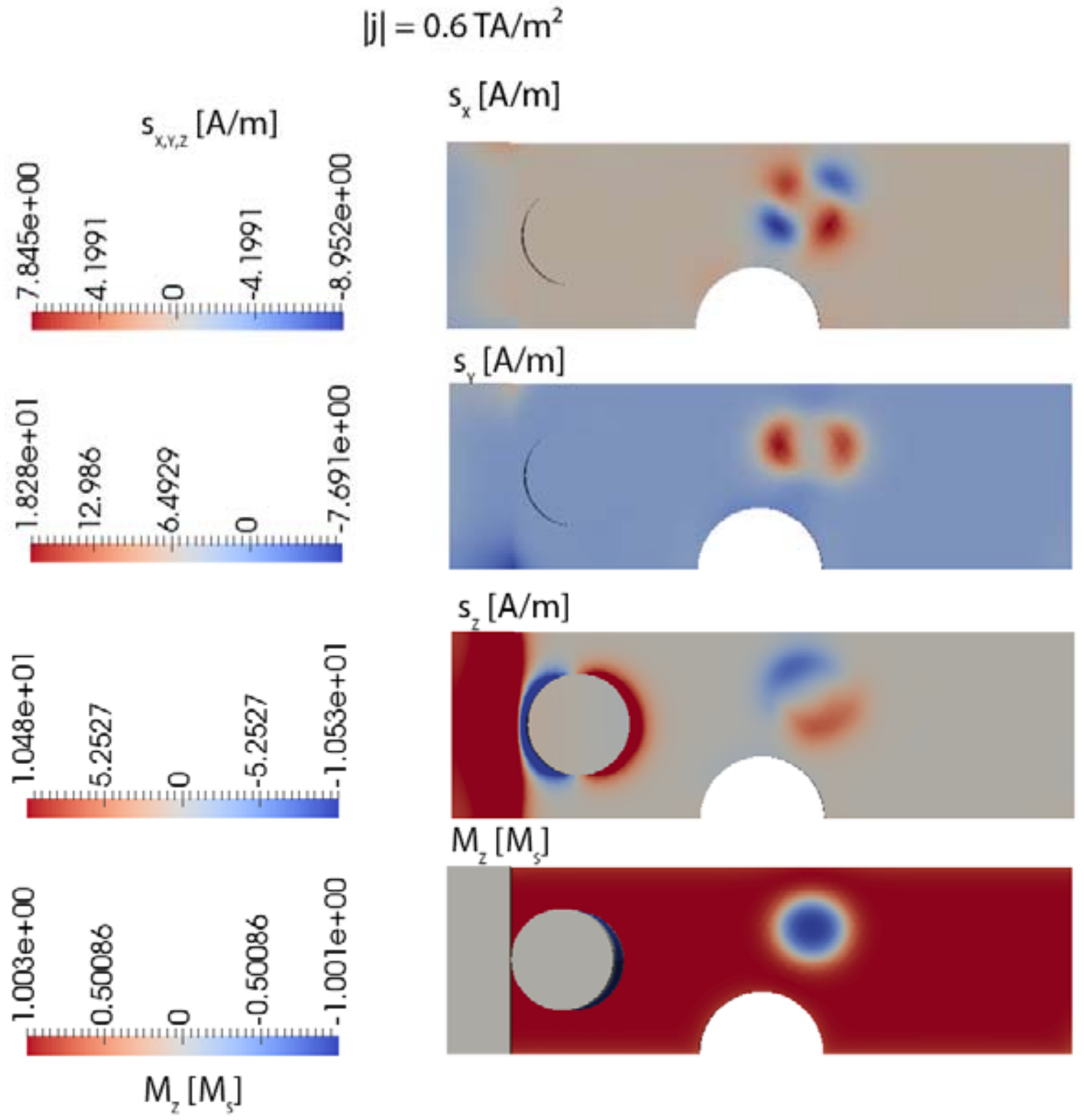

Fig. 4: Breathing modes of the simulation with $D_0 = 0$. Dimensions according to Fig. 1. The *y* and *z* component of the spin accumulation are zero ($s_y = s_z = 0$) and are not shown.

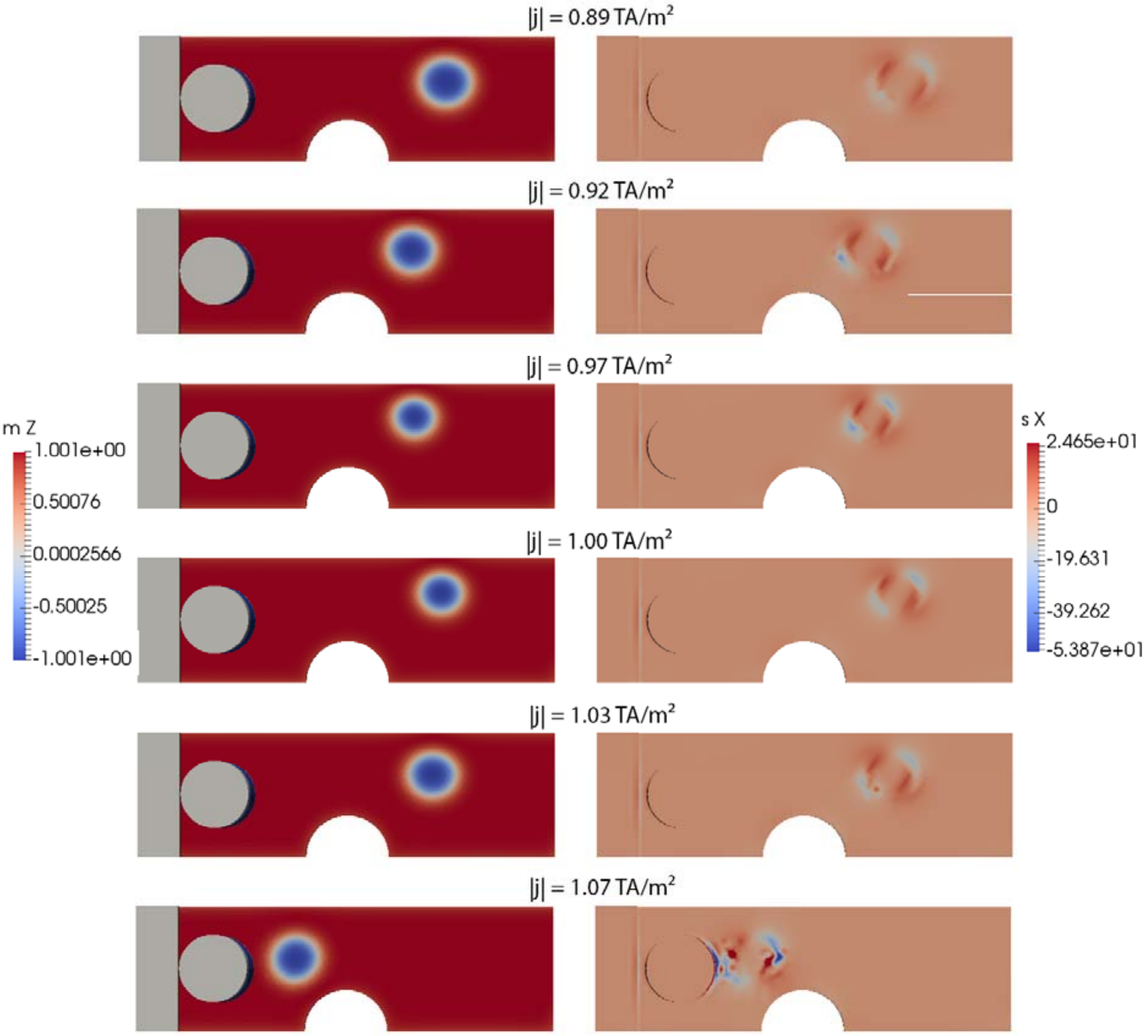

Fig. 5: States along the minimum energy path which is the most probable thermally activated reversal path. The saddle point configuration is for *i* = 10 (right column). The minimum energy states are for i=0 and i=19. The configurations correspond to a film with (right) *l* = 500 nm, *w* = 75 nm, *t* = 1.5 nm and *d* = 50nm and with (left) *l* = 450 nm, *w* = 67.5 nm, *t* = 1.35 nm and *d* = 45 nm

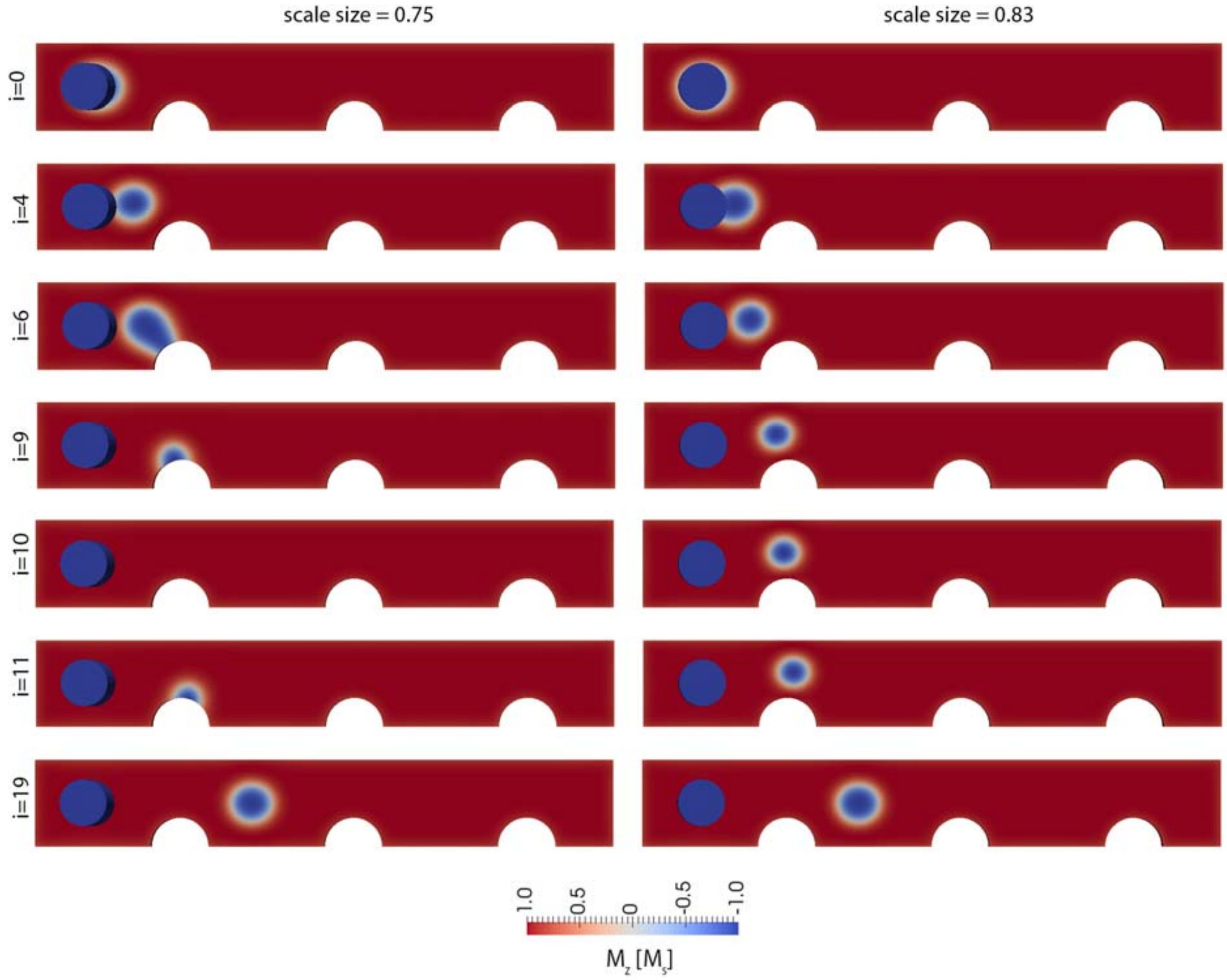

Fig. 6: Energy along the minimum energy path of the different states. The path index corresponds to the index *i* of *Fig. 5*. (black line) the simulation corresponds to the film with *l* = 500 nm, *w* = 75 nm, *t*= 1.5 nm and *d* = 50nm and with (red dashed line) *l* = 450 nm, *w* = 67.5 nm, *t*= 1.35 and *d* = 45 nm.

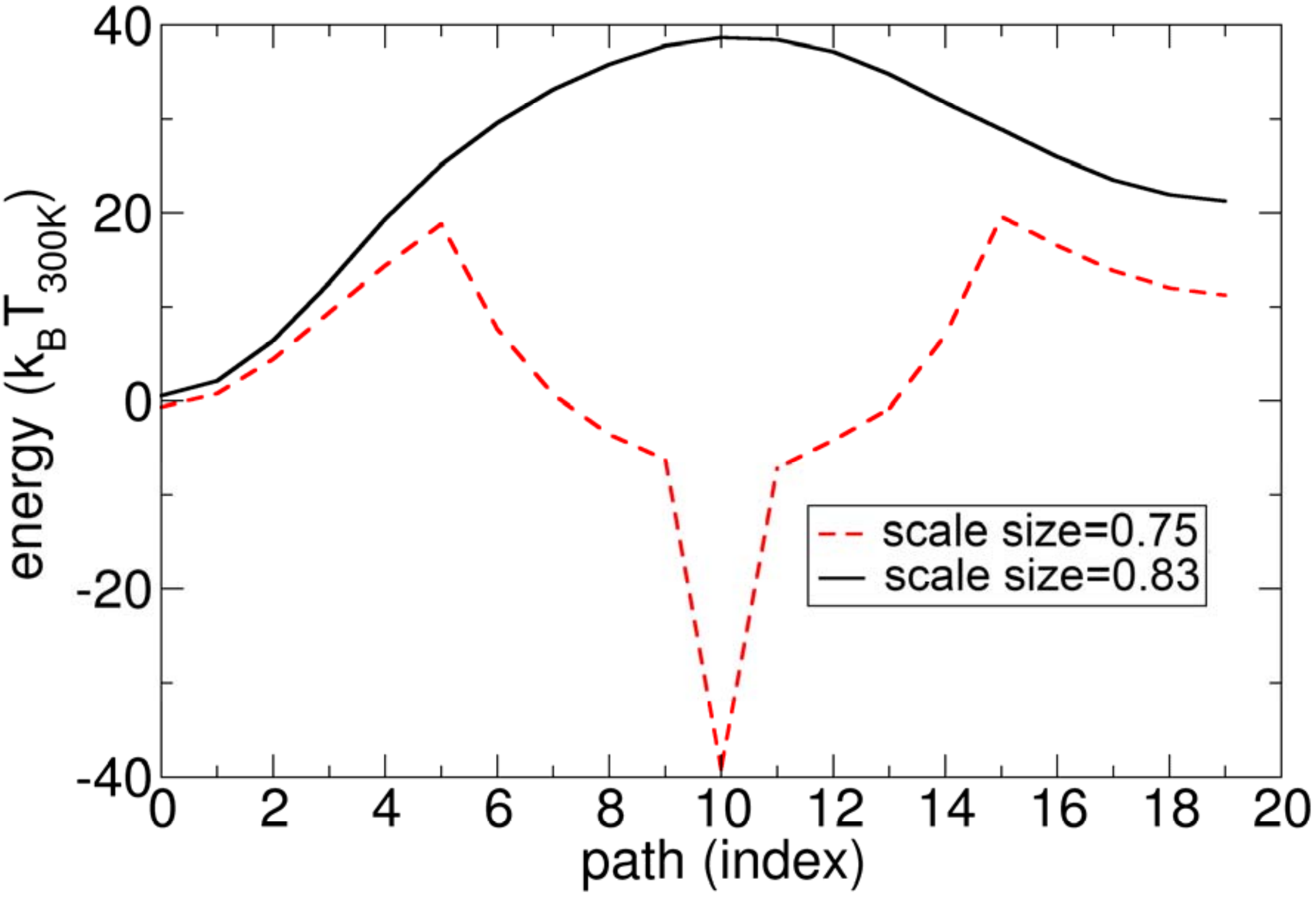

Fig. 7: Dependence of the energy barrier ΔE as function of the lateral dimension of the entire structure. The thickness is kept constant to be *t* = 1.5 nm. The simulation with *scale size* = 1 corresponds to *l* = 600 nm, *w* = 90 nm, *t* = 1.5 nm and *d* = 60nm.  For all simulations, the initial path is a motion of the skyrmion over the pinning site. For the simulations with *scale size* < 0.83 the skyrmion is annihilated at the notch.

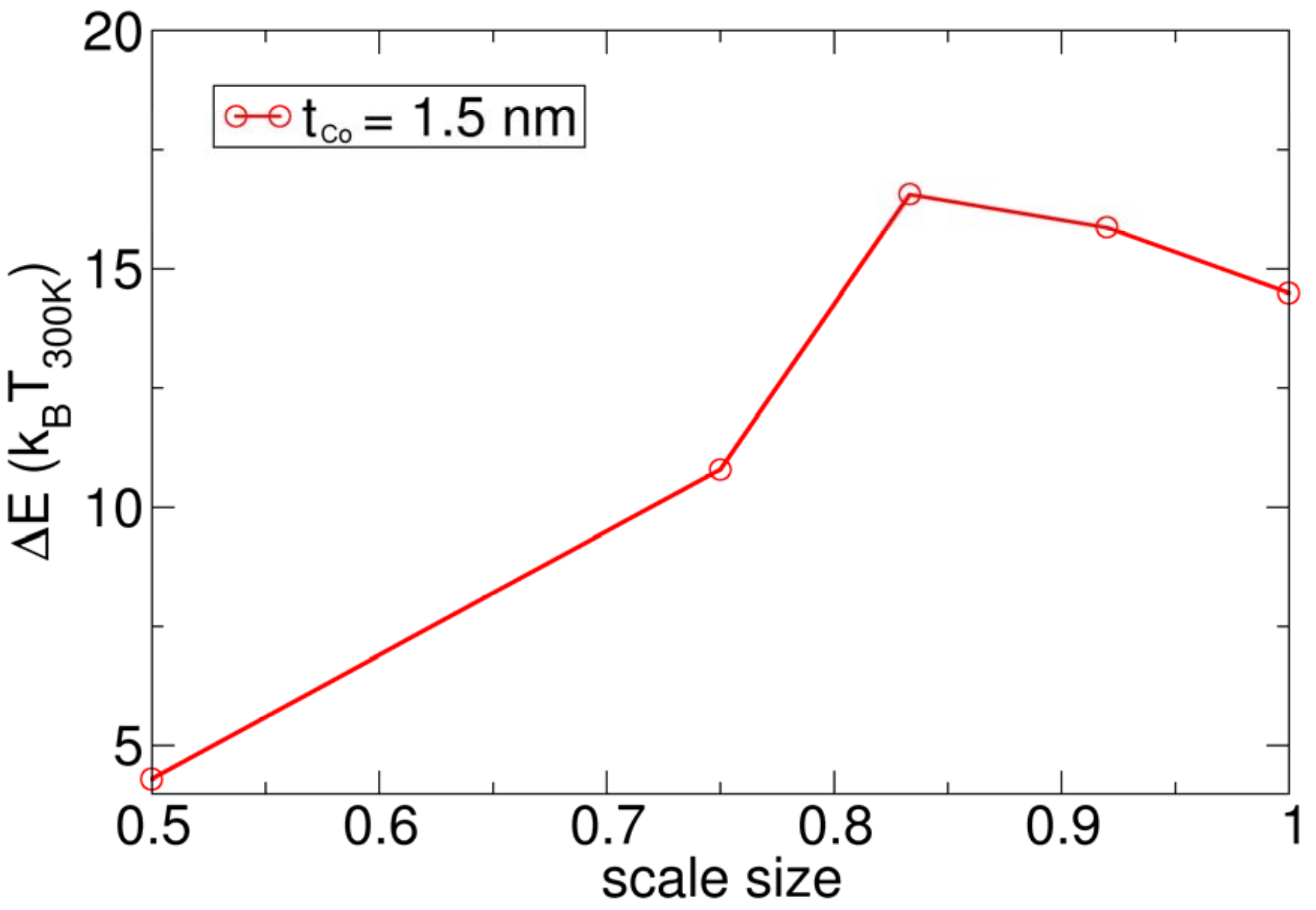

Fig. 8: Detailed view of the annihilation of a skyrmion showing that during annihilation the skyrmion changes its topological winding number without forming a Bloch point for the wire with *l* = 400 nm, *w* = 60 nm, *t* = 1.20 nm and *d* = 40 nm.

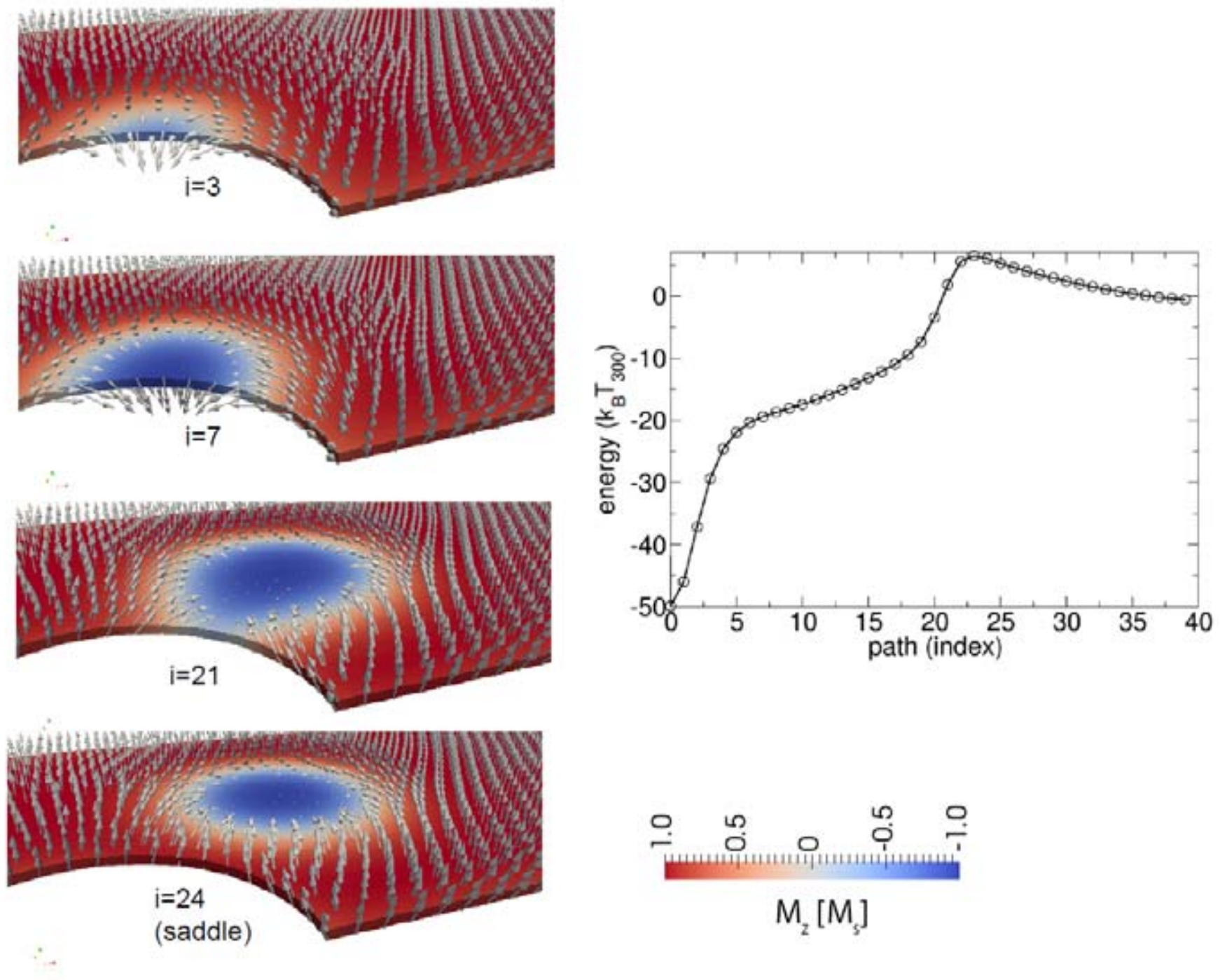

Fig. 9: Dependence of the energy barrier ΔE as function of the total thickness of Co for two structures with different lateral dimensions (circles) *l* = 500 nm, *w* = 75 nm, *d* = 50nm (rectangles) *l* = 400 nm, *w* = 60 nm, *t* = 1.2 nm and *d* = 40 nm. (red circles) the energy barrier of the path through the pinning site is calculated as function of the layer thickness. (black circles and black rectangles) the energy barrier of the path, where the skyrmion is annihilated at the boundary next to the pinning site, is calculated as function of the layer thickness. (blue circles) the energy barrier is calculated for a skyrmion which is moved out of the wire on the top flat edge without pinning sites.

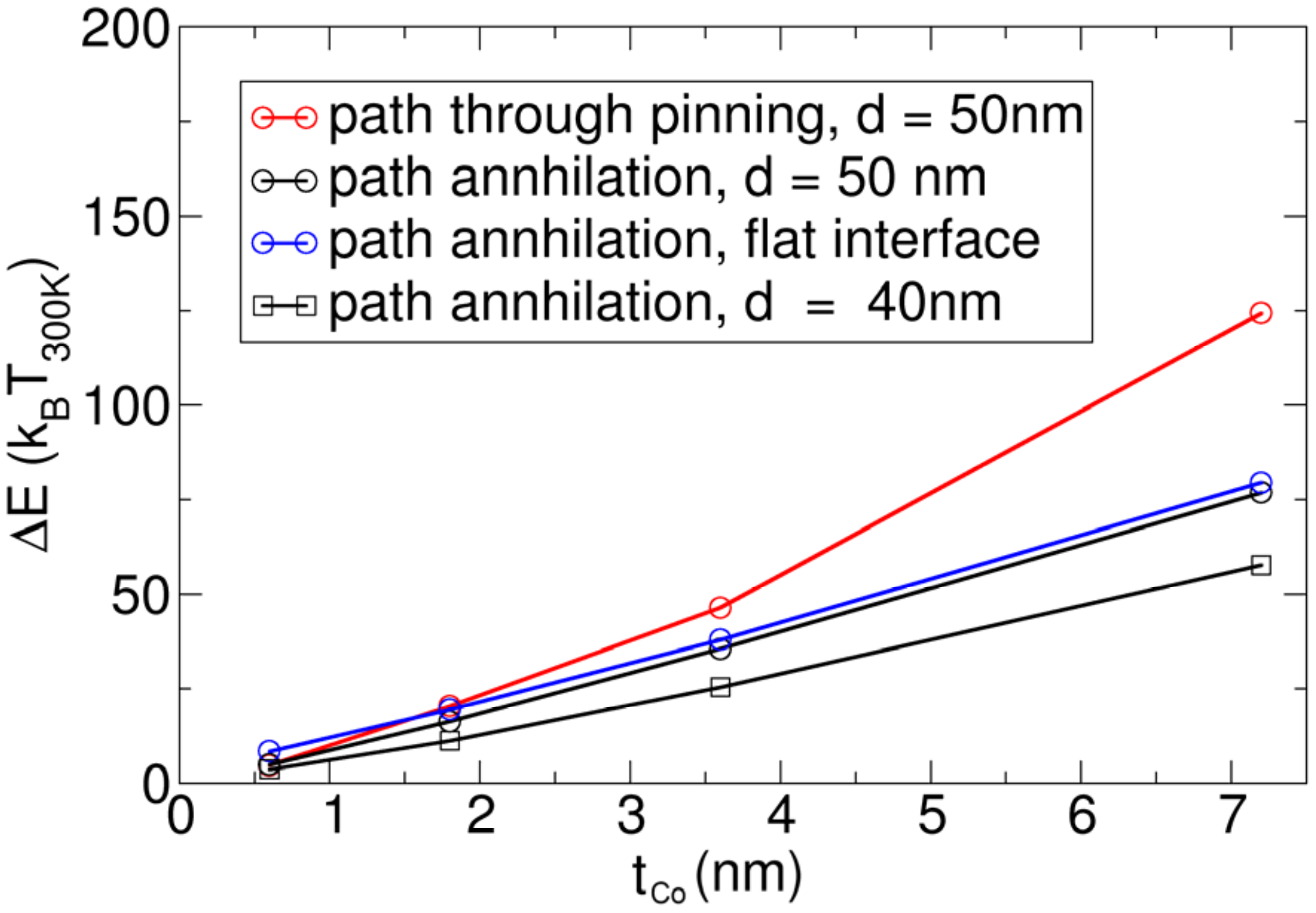

Fig. 10: Schematic illustration of the racetrack principle. (a) an applied current pulse should change the initial state (b) to the desired output (c). (d-e) two possible write errors.

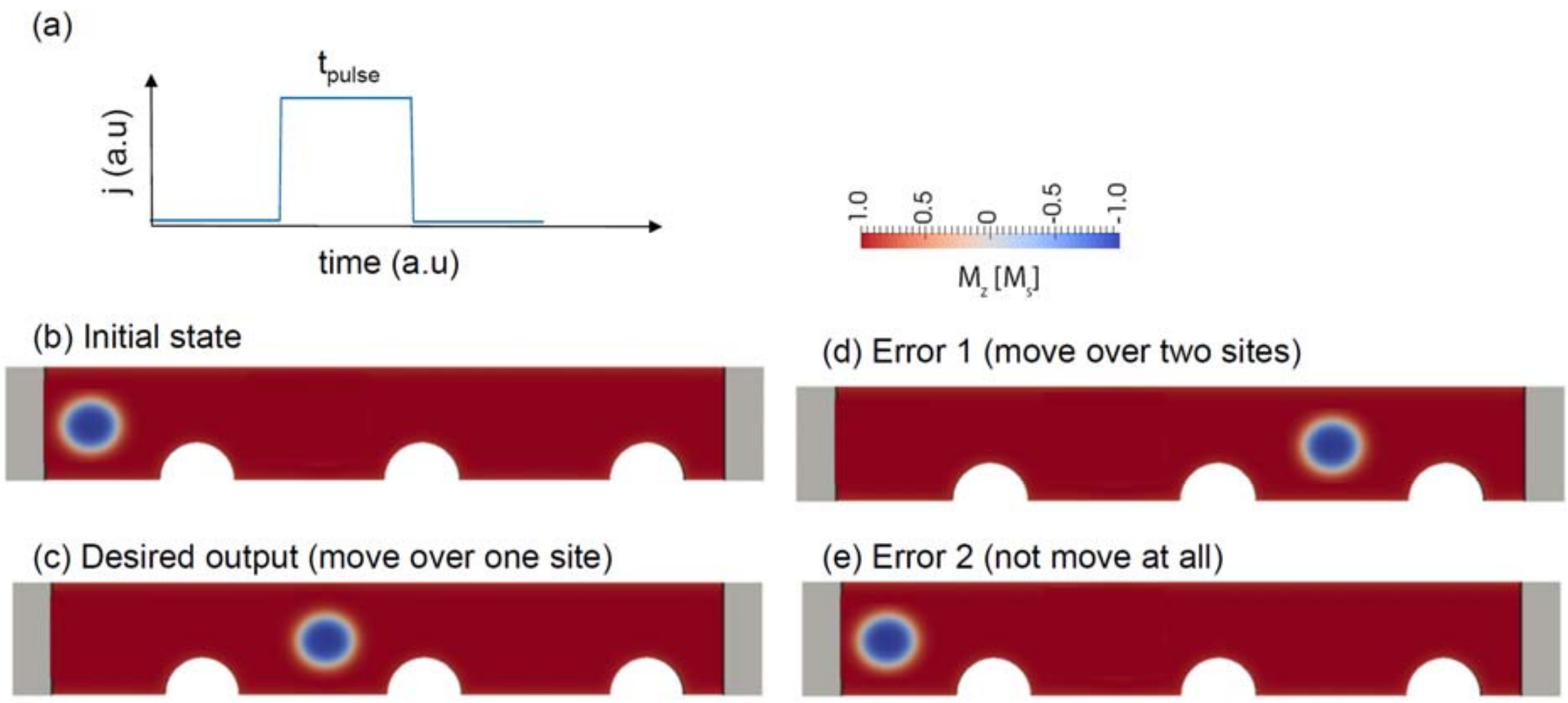

Fig. 11: Dependence of the probability for successful depinning over one pinning site and not depinning over two pinning sites as function of *N* write pulses $t_{pulse}$. The average depinning time is $t_a$ = 1ns.

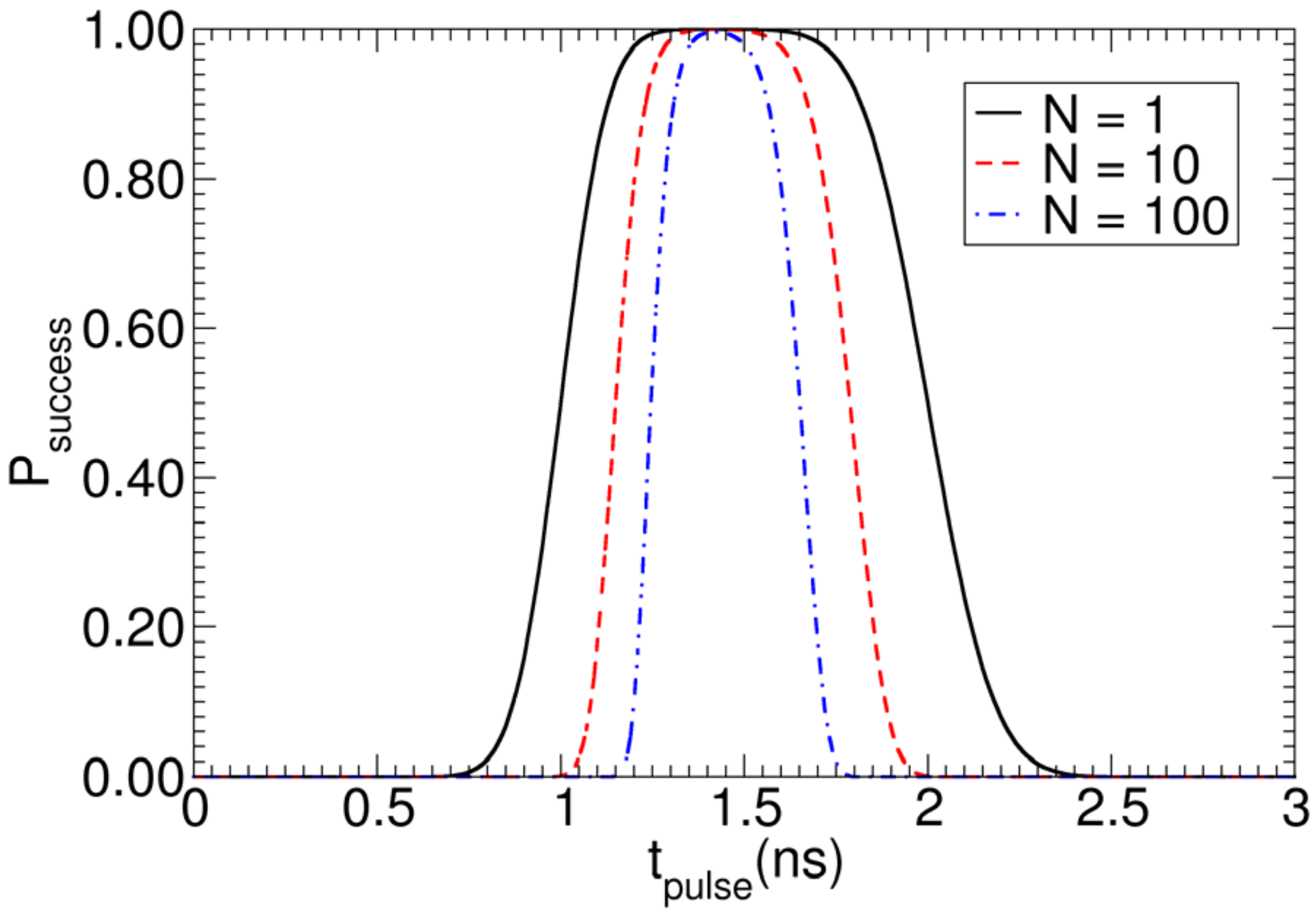

Fig. 12: Dependence of bit error rate to move a skyrmion from the input to the output for a racetrack device with $N$ bits as function of $\sigma$. For each $\sigma$ the pulse length $t_a$ was optimized to reach the highest $P_{success}$.

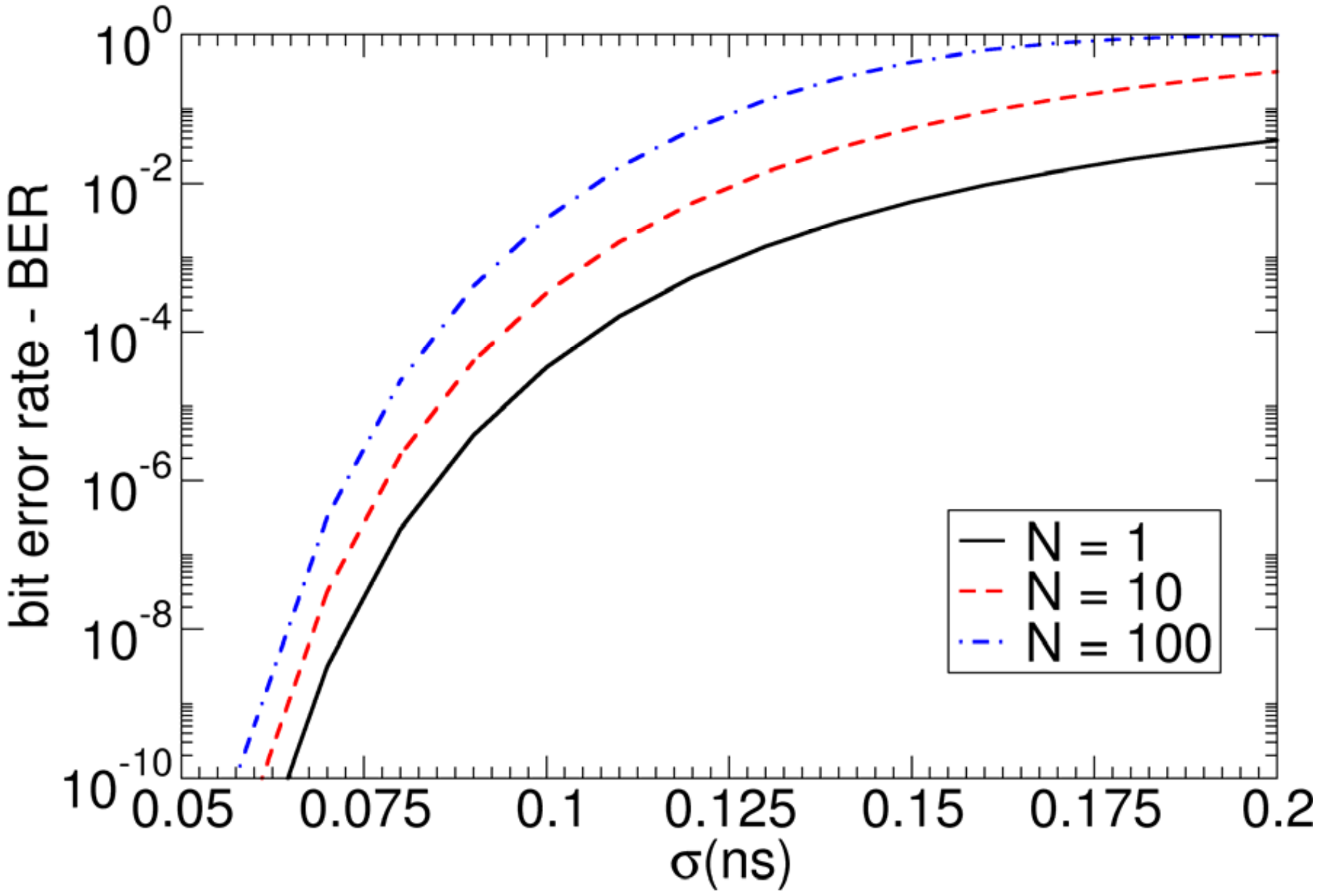

Fig. 13: Schematic representation of the (left) layered model (right) effective media model.

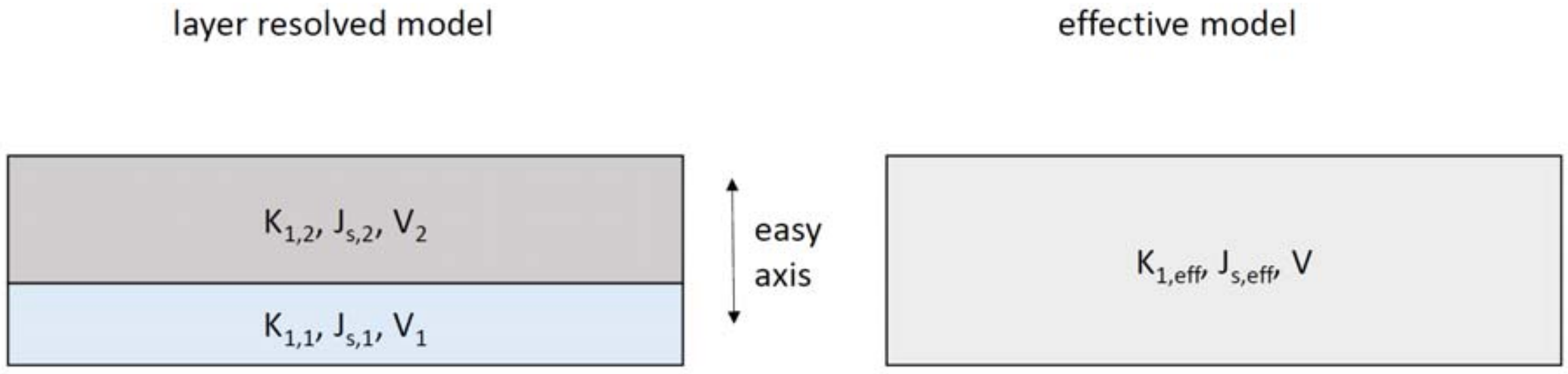

Fig. 14: Comparison of hard axis loop (field applied in-plane) of an effective media model (effective model) an model where only the Co layer are simulated and directly stacked above each other with full exchange coupling (6. Con Co layers) and an layered resolved model where between the Co layers 2 nm air is assumed since Pt and Ta are assumed non-magnetic.

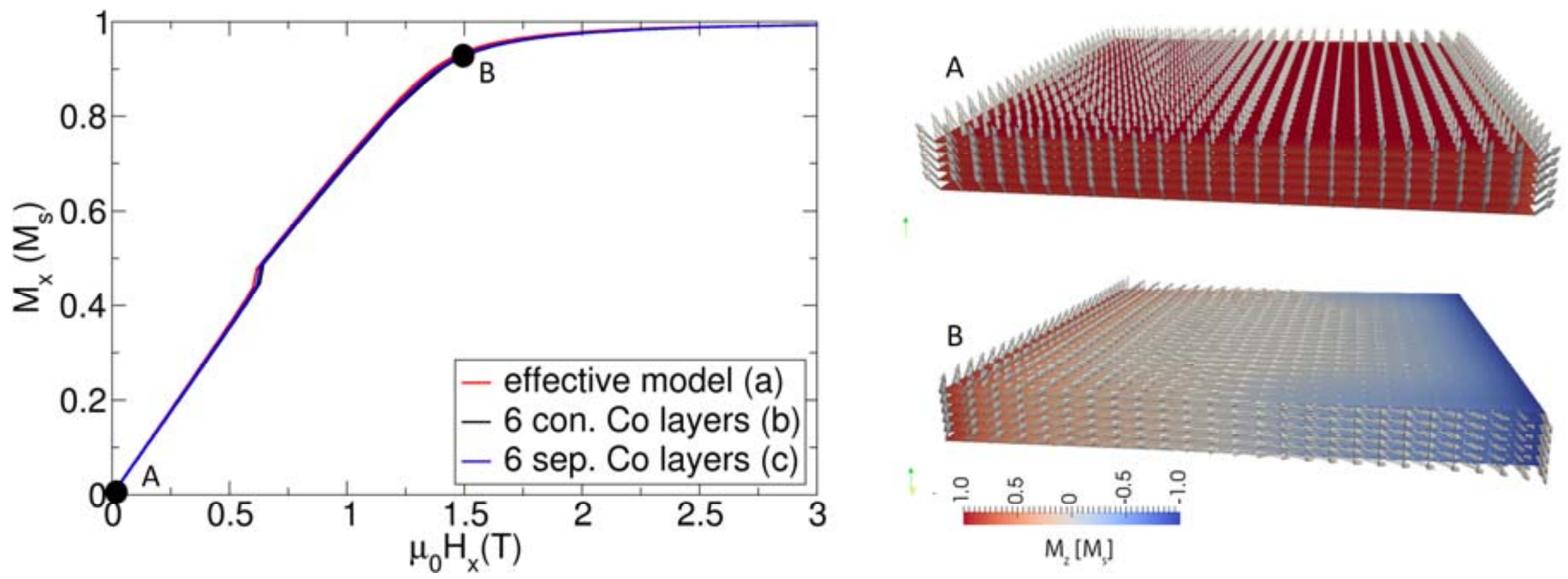

Fig. 15: Comparison of energy barrier of the three models of Fig. 14.

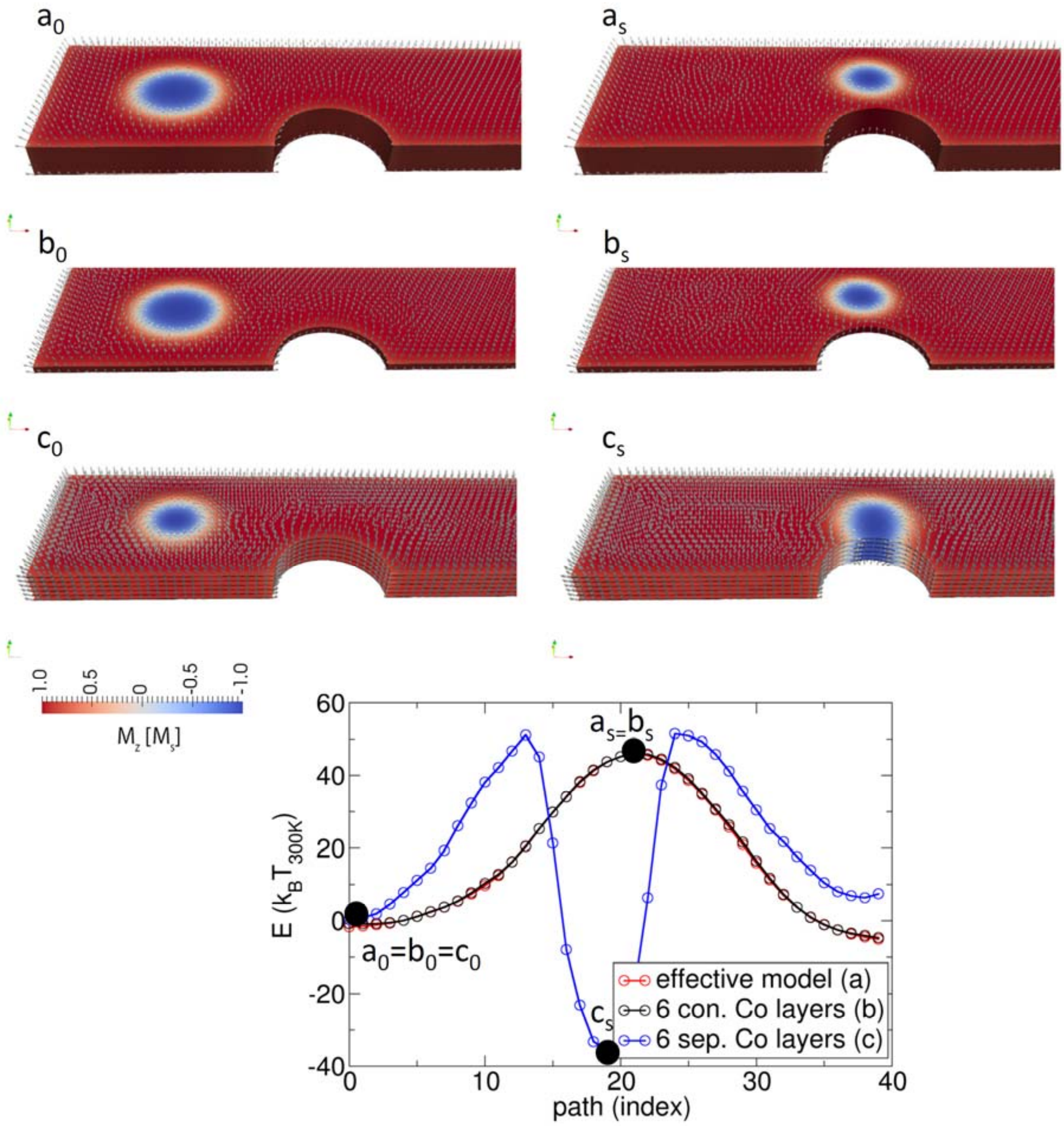

Fig. 16: Comparison of energy barrier for boundary annihilation and isotropic annihilation via a Bloch point as function of the mesh size for a system of size 90 nm x 90 nm x 0.6 nm, which represents one 0.6 nm thick Co layer (1 repetition of the layer stack).

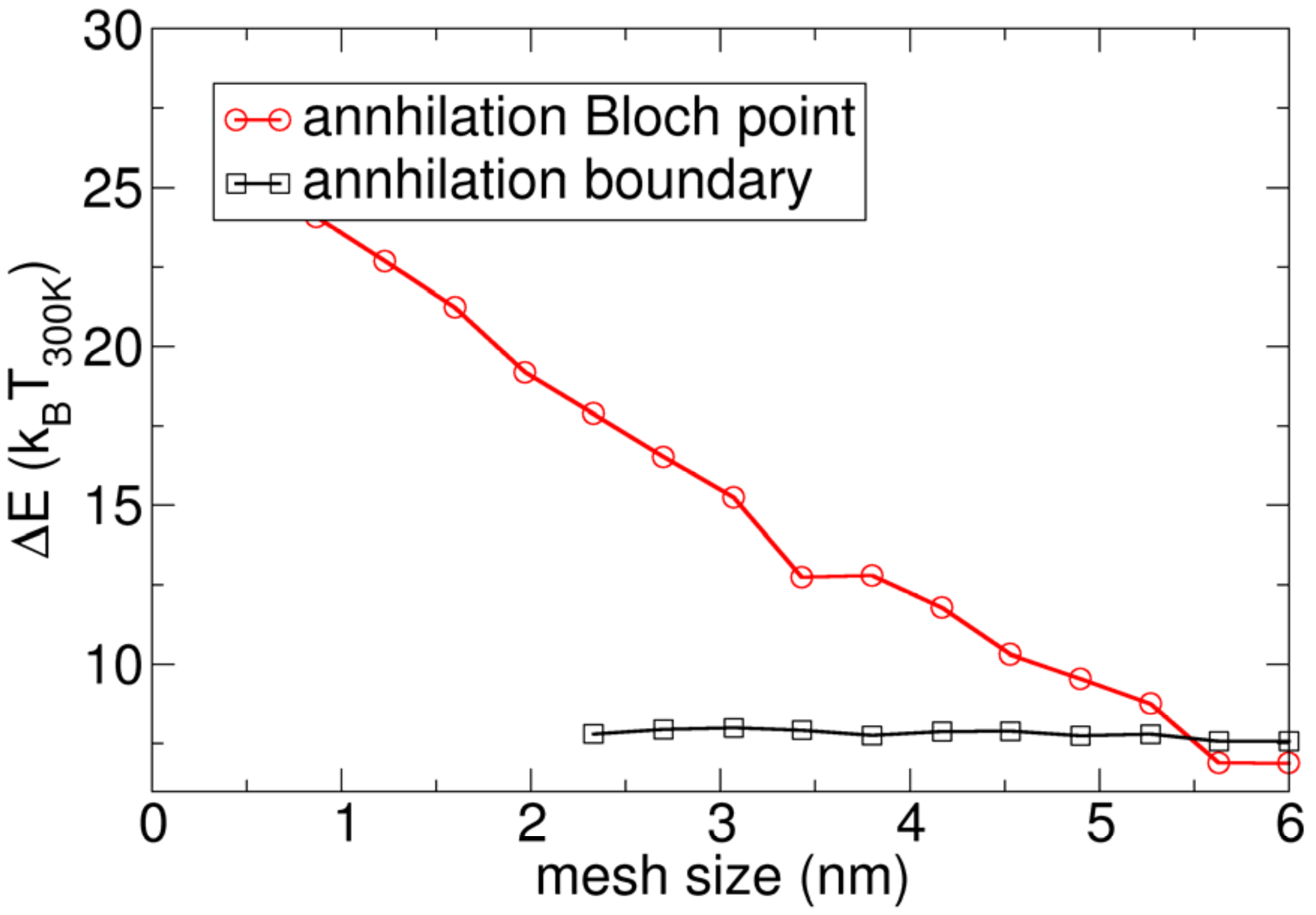

1. Bogdanov, A. N. & Rößler, U. K. Chiral Symmetry Breaking in Magnetic Thin Films and Multilayers. *Phys. Rev. Lett.* **87,** 037203 (2001).
2. Rößler, U. K., Bogdanov, A. N. & Pfleiderer, C. Spontaneous skyrmion ground states in magnetic metals. *Nature* **442,** 797–801 (2006).
3. Mühlbauer, S. *et al.* Skyrmion Lattice in a Chiral Magnet. *Science* **323,** 915–919 (2009).
4. Yu, X. Z. *et al.* Real-space observation of a two-dimensional skyrmion crystal. *Nature* **465,** 901–904 (2010).
5. Huang, S. X. & Chien, C. L. Extended Skyrmion Phase in Epitaxial FeGe ( 111 ) Thin Films. *Phys. Rev. Lett.* **108,** (2012).
6. Heinze, S. *et al.* Spontaneous atomic-scale magnetic skyrmion lattice in two dimensions. *Nat. Phys.* **7,** 713–718 (2011).
7. Zhou, Y. & Ezawa, M. A reversible conversion between a skyrmion and a domain-wall pair in a junction geometry. *Nat. Commun.* **5,** 4652 (2014).
8. Yu, X. Z. *et al.* Skyrmion flow near room temperature in an ultralow current density. *Nat. Commun.* **3,** 988 (2012).
9. Jonietz, F. *et al.* Spin Transfer Torques in MnSi at Ultralow Current Densities. *Science* **330,** 1648–1651 (2010).
10. Tomasello, R. *et al.* A strategy for the design of skyrmion racetrack memories. *Sci. Rep.* **4,** 6784 (2014).
11. Wiesendanger, R. Nanoscale magnetic skyrmions in metallic films and multilayers: a new twist for spintronics. *Nat. Rev. Mater.* **1,** 16044 (2016).
12. Fert, A., Cros, V. & Sampaio, J. Skyrmions on the track. *Nat. Nanotechnol.* **8,** 152–156 (2013).
13. Bessarab, P. F., Uzdin, V. M. & Jónsson, H. Method for finding mechanism and activation energy of magnetic transitions, applied to skyrmion and antivortex annihilation. *Comput. Phys. Commun.* **196,** 335–347 (2015).

14. Cortés-Ortuño, D. *et al.* Thermal stability and topological protection of skyrmions in nanotracks. *Sci. Rep.* **7,** 4060 (2017).

15. Stosic, D., Mulkers, J., Van Waeyenberge, B., Ludermir, T. B. & Milošević, M. V. Paths to collapse for isolated skyrmions in few-monolayer ferromagnetic films. *Phys. Rev. B* **95,** (2017).

16. Heistracher, P., Abert, C., Bruckner, F., Vogler, C. & Suess, D. GPU-Accelerated Atomistic Energy Barrier Calculations of Skyrmion Annihilations. *IEEE Trans. Magn.* 1–5 (2018). doi:10.1109/TMAG.2018.2847446

17. Rohart, S., Miltat, J. & Thiaville, A. Path to collapse for an isolated N\'eel skyrmion. *Phys. Rev. B* **93,** 214412 (2016).

18. Zhang, S., Levy, P. M. & Fert, A. Mechanisms of Spin-Polarized Current-Driven Magnetization Switching. *Phys. Rev. Lett.* **88,** 236601 (2002).

19. Abert, C. *et al.* A three-dimensional spin-diffusion model for micromagnetics. *Sci. Rep.* **5,** (2015).

20. Abert, C. *et al.* A self-consistent spin-diffusion model for micromagnetics. *Sci. Rep.* **6,** 16 (2016).

21. Woo, S. *et al.* Observation of room-temperature magnetic skyrmions and their current-driven dynamics in ultrathin metallic ferromagnets. *Nat. Mater.* **15,** 501–506 (2016).

22. Sampaio, J., Cros, V., Rohart, S., Thiaville, A. & Fert, A. Nucleation, stability and current-induced motion of isolated magnetic skyrmions in nanostructures. *Nat. Nanotechnol.* **8,** 839–844 (2013).

23. Bass, J. & Jr, W. P. P. Spin-diffusion lengths in metals and alloys, and spin-flipping at metal/metal interfaces: an experimentalist's critical review. *J. Phys. Condens. Matter* **19,** 183201 (2007).

24. Thiaville, A., Garcia, J. M., Dittrich, R., Miltat, J. & Schrefl, T. Micromagnetic study of Bloch-point-mediated vortex core reversal. *Phys. Rev. B* **67,** 094410 (2003).

25. De Lucia, A., Krüger, B., Tretiakov, O. A. & Kläui, M. Multiscale model approach for magnetization dynamics simulations. *Phys. Rev. B* **94,** 184415 (2016).

26. Troncoso, R. E. & Núñez, A. S. Thermally assisted current-driven skyrmion motion. *Phys. Rev. B* **89,** 224403 (2014).

27. Park, C. *et al.* Temperature Dependence of Critical Device Parameters in 1 Gb Perpendicular Magnetic Tunnel Junction Arrays for STT-MRAM. *IEEE Trans. Magn.* **53,** 1–4 (2017).

28. Kim, J.-V. & Yoo, M.-W. Current-driven skyrmion dynamics in disordered films. *Appl. Phys. Lett.* **110,** 132404 (2017).

29. Woo, S., Mann, M., Tan, A. J., Caretta, L. & Beach, G. S. D. Enhanced spin-orbit torques in Pt/Co/Ta heterostructures. *Appl. Phys. Lett.* **105,** 212404 (2014).

30. Djuhana, D., Supriyanto, E. & Kim, D. H. Micromagnetic Simulation of the Depinning Field Domain Wall on Symmetric Double Notch Ferromagnetic Wires. *Makara J. Sci.* 42–46 (2014). doi:10.7454/mss.v18i2.3135

31. Lepadatu, S., Vanhaverbeke, A., Atkinson, D., Allenspach, R. & Marrows, C. H. Dependence of Domain-Wall Depinning Threshold Current on Pinning Profile. *Phys. Rev. Lett.* **102,** 127203 (2009).

32. Zhang, S., Levy, P. M. & Fert, A. Mechanisms of Spin-Polarized Current-Driven Magnetization Switching. *Phys. Rev. Lett.* **88,** 236601 (2002).

33. Thiaville, A., Nakatani, Y., Miltat, J. & Suzuki, Y. Micromagnetic understanding of current-driven domain wall motion in patterned nanowires. *EPL Europhys. Lett.* **69,** 990 (2005).

34. Brown, W. F. Thermal Fluctuations of a Single-Domain Particle. *Phys. Rev.* **130,** 1677–1686 (1963).

35. Suess, D. *et al.* Calculation of coercivity of magnetic nanostructures at finite temperatures. *Phys. Rev. B* **84,** 224421 (2011).

36. Fiedler, G. *et al.* Direct calculation of the attempt frequency of magnetic structures using the finite element method. *J. Appl. Phys.* **111,** 093917 (2012).

37. Vogler, C. *et al.* Simulating rare switching events of magnetic nanostructures with forward flux sampling. *Phys. Rev. B - Condens. Matter Mater. Phys.* **88,** (2013).

38. Bessarab, P. F. *et al.* Lifetime of racetrack skyrmions. *Sci. Rep.* **8,** 3433 (2018).

39. Desplat, L., Suess, D., Kim, J.-V. & Stamps, R. L. Thermal stability of metastable magnetic skyrmions: Entropic narrowing and significance of internal eigenmodes. *ArXiv180206744 Cond-Mat* (2018).

40. E, W., Ren, W. & Vanden-Eijnden, E. String method for the study of rare events. *Phys. Rev. B* **66,** 052301 (2002).

41. E, W., Ren, W. & Vanden-Eijnden, E. Simplified and improved string method for computing the minimum energy paths in barrier-crossing events. *J. Chem. Phys.* **126,** 164103 (2007).

42. Cortés-Ortuño, D. *et al.* Thermal stability and topological protection of skyrmions in nanotracks. *ArXiv161107079 Cond-Mat* (2016).

43. Braun, H.-B. Topological effects in nanomagnetism: from superparamagnetism to chiral quantum solitons. *Adv. Phys.* **61,** 1–116 (2012).

44. Fert, A., Reyren, N. & Cros, V. Magnetic skyrmions: advances in physics and potential applications. *Nat. Rev. Mater.* **2,** 17031 (2017).

45. APS -APS March Meeting 2016 - Event - Skyrmions in thin-film multilayers with interfacially-induced Dzyaloshinskii-Moriya interaction observed by MFM. in *Bulletin of the American Physical Society* **Volume 61, Number 2,** (American Physical Society).

46. Liu, Z. Y. *et al.* Oscillatory antiferromagnetic interlayer coupling in \$\mathrm{Co}(4\phantom{\rule{0.3em}{0ex}}\mathrm{\AA{}})⁄\mathrm{Pt}({t}_{\mathrm{Pt}}\phantom{\rule{0.3em}{0ex}}\mathrm{\AA{}})⁄[\mathrm{Co}(4\phantom{\rule{0.3em}{0ex}}\mathrm{\AA{}})⁄\mathrm{Pt}(6\phantom{\rule{0.3em}{0ex}}\mathrm{\AA{}})⁄\mathrm{Co}(4\phantom{\rule{0.3em}{0ex}}\mathrm{\AA{}})]⁄\mathrm{Ni}\mathrm{O}(20\phantom{\rule{0.3em}{0ex}}\mathrm{\AA{}})\$ multilayers with perpendicular anisotropy. *Phys. Rev. B* **77,** 012409 (2008).

47. Suess, D., Vogler, C., Bruckner, F., Sepehri-Amin, H. & Abert, C. Significant reduction of critical currents in MRAM designs using dual free layer with dynamical perpendicular and in-plane anisotropy. *ArXiv Prepr. ArXiv170200996* (2017).

48. Sun, J. Z. *et al.* Spin-torque switching efficiency in CoFeB-MgO based tunnel junctions. *Phys. Rev. B* **88,** 104426 (2013).

49. Im, M.-Y., Bocklage, L., Fischer, P. & Meier, G. Direct Observation of Stochastic Domain-Wall Depinning in Magnetic Nanowires. *Phys. Rev. Lett.* **102,** 147204 (2009).

50. Burrowes, C. *et al.* Non-adiabatic spin-torques in narrow magnetic domain walls. *Nat. Phys.* **6,** 17–21 (2010).

51. Briones, J. *et al.* Stochastic and complex depinning dynamics of magnetic domain walls. *Phys. Rev. B* **83,** 060401 (2011).

52. Mihai, A. P. *et al.* Stochastic domain-wall depinning under current in FePt spin valves and single layers. *Phys. Rev. B* **84,** 014411 (2011).

53. Spin-transfer pulse switching: From the dynamic to the thermally activated regime. *Appl. Phys. Lett.* **97,** 262502 (2010).

54. Litzius, K. *et al.* Skyrmion Hall effect revealed by direct time-resolved X-ray microscopy. *Nat. Phys.* **13,** 170–175 (2017).

55. Lee, Y.-J., Wang, P.-K. & Jan, G. MRAM write pulses to dissipate intermediate state domains. (2016).

56. Müller, J. Magnetic skyrmions on a two-lane racetrack. *New J. Phys.* **19,** 025002 (2017).

57. Suess, D., Vogler, C., Bruckner, F. & Abert, C. A repulsive skyrmion chain as guiding track for a race track memory. *ArXiv170706925 Cond-Mat* (2017).

58. Ju, G. *et al.* High Density Heat-Assisted Magnetic Recording Media and Advanced Characterization #x2014;Progress and Challenges. *IEEE Trans. Magn.* **51,** 1–9 (2015).

59. Wiesendanger, R. Nanoscale magnetic skyrmions in metallic films and multilayers: a new twist for spintronics. *Nat. Rev. Mater.* **1,** 16044 (2016).

60. Mittal, S. A Survey of Techniques for Architecting Processor Components Using Domain-Wall Memory. *J Emerg Technol Comput Syst* **13,** 29:1–29:25 (2016).

61. Suess, D., Fidler, J. & Schrefl, T. chapter 2 Micromagnetic Simulation of Magnetic Materials. *Handb. Magn. Mater.* **16,** 41–125 (2006).

62. Thiaville, A., Rohart, S., Jué, É., Cros, V. & Fert, A. Dynamics of Dzyaloshinskii domain walls in ultrathin magnetic films. *EPL Europhys. Lett.* **100,** 57002 (2012).

63. Dittrich, R. *et al.* A path method for finding energy barriers and minimum energy paths in complex micromagnetic systems. *J. Magn. Magn. Mater.* **250,** L12–L19 (2002).

64. Henkelman, G., Uberuaga, B. P. & Jónsson, H. A climbing image nudged elastic band method for finding saddle points and minimum energy paths. *J. Chem. Phys.* **113,** 9901–9904 (2000).

65. Abert, C., Exl, L., Bruckner, F., Drews, A. & Suess, D. Magnum.fe: A micromagnetic finite-element simulation code based on FEniCS. *J. Magn. Magn. Mater.* **345,** 29–35 (2013).

66. Heistracher, P. T. *Master Thesis, Atomistic spin dynamics, http://repositum.tuwien.ac.at/obvutwhs/content/titleinfo/2281383*. (Wien, 2017).